\documentclass[10pt]{article}
\usepackage[utf8]{inputenc}
\usepackage[margin=1in]{geometry}
\usepackage{times}
\usepackage{amsfonts,amsmath,amssymb,setspace,xcolor, graphicx}
\usepackage{soul,overpic,mathtools}
\usepackage[export]{adjustbox}
\usepackage{float}
\usepackage{array}
\usepackage{multirow}
\usepackage{float}
\usepackage[utf8]{inputenc}
\usepackage{enumitem}
\usepackage{comment}
\usepackage{makecell}
\usepackage{setspace}
\usepackage[toc,page]{appendix}

\usepackage[
sorting=none,
style=numeric-comp,
]{biblatex}
\usepackage{bbold}
\usepackage{lineno}
\title{Ensuring resilience to extreme weather events increases the ambition of mitigation scenarios on solar power and storage uptake: a study on the Italian power system}
\author{Alice Di Bella$^{1,3,4,*}$, Francesco Colelli$^{2,3,4}$}
\date{%
$^1$ Department of Electronics, Information and Bioengineering, Politecnico di Milano, Milano (MI), Italy \\[2ex]
$^2$ Department of Economics, Ca’ Foscari University of Venice, 30121 Venice, Italy \\[2ex]
$^3$ CMCC Foundation - Euro-Mediterranean Center on Climate Change, Italy \\[2ex]
$^4$ RFF-CMCC European Institute on Economics and the Environment, Italy \\[2ex]
\footnotesize *Corresponding author. \textit{e-mail address:} alice.dibella@cmcc.it
\bigskip
    }

\nolinenumbers
\begin{document}

\maketitle

\begin{abstract}

This study explores compounding impacts of climate change on power system's load and generation, emphasising the need to integrate adaptation and mitigation strategies into investment planning. We combine existing and novel empirical evidence to model impacts on: i) air-conditioning demand; ii) thermal power outages; iii) hydro-power generation shortages. Using a power dispatch and capacity expansion model, we analyse the Italian power system's response to these climate impacts in 2030, integrating mitigation targets and optimising for cost-efficiency at an hourly resolution. We outline different meteorological scenarios to explore the impacts of both average climatic changes and the intensification of extreme weather events. We find that addressing extreme weather in power system planning will require an extra 5-8 GW of  photovoltaic  (PV) capacity, on top of the 50 GW of the additional solar PV capacity required by the mitigation target alone. Despite the higher initial investments, we find that the adoption of renewable technologies, especially PV, alleviates the power system's vulnerability to climate change and extreme weather events. Furthermore, enhancing short-term storage with lithium-ion batteries is crucial to counterbalance the reduced availability of dispatchable hydro generation.

\end{abstract}


\thispagestyle{empty}
\clearpage
\doublespacing
\nolinenumbers

\section{Introduction}
% Outlining the general problem

The complex and urgent challenge posed by climate change, including more frequent and severe windstorms, heavy precipitation, droughts, and wildfires, can significantly increase risks to power infrastructure and energy systems \cite{yalew2020impacts}. %Several are increasingly acknowledging their responsibility and are committing to address climate change through international treaties and ambitious targets \cite{noauthor_paris_2016}.
At the same time, nations are increasingly committing to fundamentally reshape energy infrastructures, transitioning them towards low-carbon solutions to mitigate the adverse effects of climate change \cite{noauthor_paris_2016}. Nevertheless, the impacts of climate change are already evident in daily life, with projections indicating that these effects will intensify \cite{noauthor_summary_nodate}. As a result, 
it is imperative to design and plan future energy and power systems with a dual focus: not only on mitigating climate change but also on ensuring adaptation to the impacts that will arise.
% Specific problem: climate change on power sector
% mitigation + adaptation interplay
The need for planning with a focus on both climate change mitigation and adaptation is particularly crucial when considering the electricity sector.% This is due to two key factors that make power system planning especially complex. 
Firstly, future power infrastructure will increasingly depend on variable renewable energy sources (VREs), which are far more weather-dependent than traditional fossil-based thermal plants \cite{berrill_environmental_2016}. Secondly, a widely recommended strategy for decarbonising other sectors of the economy is the electrification of final demand \cite{noauthor_electrification_nodate}. This shift will require significant increases in electricity production, which may be more susceptible to weather-related risks due to the higher penetration of VREs. Consequently, careful planning is essential to ensure that future power systems are both sustainable and resilient in the face of these challenges.
% Geographical focus
In the case of the European Union (EU), power systems are already  undergoing significant transformations, with nearly one-third of the region's electricity now generated from renewable sources such as hydropower, solar, and onshore wind \cite{noauthor_european_2023}.  As electrification progresses and the integration of RES deepens, European energy systems are becoming increasingly dependent on weather conditions. 

% Describe the impacts of climate change on the power sector - DEMAND SIDE
Climate change impacts every stage of the electricity generation process: supply side, transmission and distribution networks, and load patterns. On the demand side, elevated temperatures can lead to increased electricity consumption for air conditioning during the summer, while warmer winters may reduce demand for heating: these changes affect not only the overall electricity demand but also result in higher load peaks and alterations in the power load profile \cite{colelli_air-conditioning_2023,hamlet_effects_2010,taseska_evaluation_2012,franco_climate_2008,auffhammer_climate_2017,parkpoom_analyzing_2008, ahmed_climate_2012}. As populations adapt to climate change, analyses on the influence of demand-side shocks on power systems are crucial for planning resilient systems. Some studies consider aggregated timespans, seasonal parameters or daily peaks for the electricity load \cite{ahmed_climate_2012,wenz2017north,eskeland_electricity_2010}, while others reach an hourly resolution for temperatures and power demand \cite{parkpoom_analyzing_2008,crowley_hourly_nodate}. Having estimates at a fine detail appears critical to enhance the resilience of the power sector, which could suffer during hours of high demand and low availability of supply \cite{bartos2015impacts}. Most of the literature has provided evidence of the short-term, intensive margin adjustments, examining how immediate changes in temperature influence electricity demand on a day-to-day or hour-by-hour basis \cite{wenz2017north,auffhammer2017climate,fonseca2019seasonal,romitti2022heterogeneous}, disregarding instead long-run extensive margin adjustments. Over the past two decades, AC usage has increased rapidly in both developed and developing regions \cite{IEA2018}, highlighting the critical need for novel analyses to better understand the implications of AC adoption on power system planning \cite{colelli_air-conditioning_2023}. %A critical gap in quantifying the impacts of climate change lies in understanding how the increased frequency and intensity of temperature extremes will amplify power demand to levels that exceed current system capacities, particularly when considering the endogenous adoption of residential air conditioning \cite{colelli_air-conditioning_2023}. 

% Describe the impacts of climate change on the power sector - SUPPLY SIDE
As previously introduced, climate change also has significant impacts on the generation side of the power infrastructure. Higher air and water temperatures can reduce the cooling efficiency of thermal power plants, leading to reduced power output. Coal and nuclear plants, which operate using steam-turbine processes, can experience significant operational challenges during droughts. Variations in stream flow levels and elevated temperatures can substantially impact the availability of cooling water required for these plants to function at full capacity \cite{bartos2016impacts}. Gas-fired power plants, operating though combustion-turbine processes that require little or no water for cooling, can be affected by a reduction in the efficiency of turbines due to extreme ambient temperatures, ultimately leading to power output reductions \cite{bartos2015impacts}. Recent work has underscored that climate change has already increased curtailment of thermal power plants \cite{coffel2021thermal}.

Climate change can increase the frequency of prolonged droughts, which can reduce water availability for hydropower. A large body of literature has focused on how water scarcity due to climate change can undermine power generation of hydroelectric dams \cite{van2016multi,SOLAUN2019,LU2020,hamududu_assessing_2016, turner_climate_2017, van_vliet_vulnerability_2012}. van Vliet et al. observe that climate change is projected to lead to significant regional variations in hydropower potential: some areas are expected to experience substantial increases in potential, while others may face considerable decreases, such as the United States and southern Europe \cite{van2016multi}. Turner et al. corroborate these findings, highlighting the pronounced regional variability in hydropower impacts and emphasising the considerable investment required to adapt to anticipated changes in water availability \cite{turner_climate_2017}. Climate change also significantly impacts renewable power generation. Heatwaves and high temperatures can reduce the efficiency of solar panels, while altered wind patterns can affect wind turbine output \cite{yalew2020impacts, jerez_impact_2015}. Evaluating these effects is complex, and there is limited consensus on the magnitude and direction of climate-induced impacts on variable RES, particularly when addressing issues at the country level \cite{cronin_climate_2018, bloomfield_quantifying_2021,hu_implications_2023, tobin_climate_2016, jerez_impact_2015, liu_climate_2023, gernaat_climate_2021}.

Finally, weather-related impacts on transmission and distribution systems are expected to be amplified by climate change. Increased frequency and severity of extreme weather events, such as storms and high winds, can cause physical damage to infrastructure, leading to outages and disruptions \cite{panteli_influence_2015}. Additionally, higher temperatures can affect the efficiency and capacity of transmission lines, increasing the risk of overheating and reducing performances \cite{hu_implications_2023, cohen_multi-model_2022, stankovski_power_2023}. Transmission capacity, crucial for the integration of the future electricity grid, might be significantly reduced during summertime due to increased peak temperatures \cite{bartos2015impacts}. %Also, Bartos and Chester \cite{bartos2016impacts} estimated that almost half of current power stations in the US might be vulnerable to climate before 2050.

To provide policymakers with strategies for enhancing the resilience of power systems, integrating climate change impacts on both supply and demand into a comprehensive modelling framework is essential. Weather and climate effects can disrupt electricity operations, leading to increased costs, load shedding, or outages if demand exceeds forecasts or power capacities are compromised. Only a few studies incorporate high-frequency supply and demand forecasts under climate change, enabling an evaluation of effective responses. 
Tobin et al. \cite{tobin_vulnerabilities_2018} assess climate impacts on European electricity production and suggest increasing resilient RES like wind and solar. 
Bloomfield et al. \cite{bloomfield_quantifying_2021} highlight substantial climate-induced variability in Europe’s energy balance by 2050, stressing the need for better integration of climate uncertainty in planning.

Optimisation models are traditionally employed to develop strategies for power capacity planning, especially when incorporating future mitigation goals \cite{pfenninger_energy_2014}, while the integration of empirically-estimated climate change impacts in energy models is more limited (country-level studies include
%. Some studies model climate impacts in Integrated Assessment Models (IAM) or macro economic models, offering a broad view over the links and consequences on socio economic variables. Arango-Aramburo et al. 
\cite{arango-aramburo_climate_2019} 
%chose the regional scope of the Columbian electricity system, focusing on hydropower, a fundamental resource for power supply in the country. The research employs partial and general equilibrium models and finds that climate change will significantly reduce Colombia's hydropower capacity by 2050, necessitating diversification into renewables, demand reduction, or fossil fuel expansion with carbon capture to meet future energy needs. De Lucena et al. 
and \cite{de_lucena_least-cost_2010}).
%study the impacts on the Brazilian electricity through a coupling of the analysis tool MAED (Model for Analysis of Energy Demand) and the integrated assessment model MESSAGE (Model for Energy Supply Strategy Alternatives and their General Environmental impact). Colelli et al. \cite{colelli2022increased} offer recent projections from IAMs, indicating that the energy required to adapt to higher temperatures can result in significant additional generation capacity and larger emissions of greenhouse gases and local air pollutants, in turn affecting the ambition of mitigation policies. 
IAM-based projections provide valuable insights; however, their findings tend to be aggregated in both space and time due to the broad nature of aggregated energy demand \cite{colelli2022increased}. In contrast, high-resolution bottom-up power system models offer more detailed analysis, making them better suited for understanding the complexities of the interplay between mitigation and adaptation within the power system. Bottom-up models might advance our vision on the hourly and local consequences of future climate, since they generally have a larger spatial and temporal resolution compared to IAMs \cite{gernaat_climate_2021}. In a few cases, regional versions of leading IAMs have been coupled with power-system models with the aim of assessing climate change impacts on peak load and generation (\cite{zhou_modeling_2014}, \cite{dowling_impact_2013},\cite{handayani_seeking_2020}). 
%conducted an analysis of the European energy system using the Prospective Outlook on Long-term Energy Systems (POLES) model, taking into account impacts on power demand and generation, with results indicating a stronger effect on the demand side. Teotonio et al. \cite{teotonio_assessing_2017} assessed the possible unavailability of hydropower in the projected Portuguese power generation with the model TIMES\_PT, observing that hydropower production may decrease by 41\% in 2050. Handayani et al. \cite{handayani_seeking_2020} study the interrelations of climate change impacts and carbon dioxide mitigation policies for the Java-Bali electricity grid through the Long-range Energy Alternative Planning system (LEAP) model. Their research includes mitigation strategies in the picture, while it does not rely on empirical methodologies for the assessment of the climate scenarios. In the case of joint efforts for mitigation and adaptation, outcomes reveal a notable escalation in installed capacities alongside a concurrent rise in investment needs. 
To the best of our knowledge, Handayani et al. is one of the few research articles that thoroughly explores the intricate relationship between optimising mitigation and adaptation strategies in power sector planning \cite{handayani_seeking_2020}.

%The investigation of extreme weather events under climate change scenarios has received less attention compared to gradual changes, likely because of great uncertainty in this area of climate science \cite{cronin_climate_2018}. %Moreover research linking empirically evaluated climate change impacts and mitigation strategies in power system models is lacking \cite{yalew_impacts_2020}. 
%Previous studies have addressed the issues of possible effects of changing climate with a European scope \cite{van_vliet_vulnerability_2012, dowling_impact_2013,zachariadis_effect_2014, tobin_vulnerabilities_2018}, but missing the mitigation perspective, while research on transition pathways does not offer a comprehensive view of the resilience of the power sector \cite{odenberger_pathways_2010,plesmann_how_2017, neumann_potential_2023}.  

In this context, this paper makes two significant contributions to both academic literature and power system investment planning. First, it stands out as one of the few studies that simultaneously addresses decarbonization goals and the impacts of climate change on various power generation technologies and electricity demand while modelling power system expansions. Unlike existing research that typically focuses on long-term pathways (e.g.\cite{handayani_seeking_2020}), this paper examines the operation and optimal capacity requirements for a specific year. This approach provides high temporal (hourly) and spatial (market zones) resolution, which is crucial given that climate change effects can be highly localised and may vary significantly at different times of the day. As a second contribution, by delineating various scenarios, this paper addresses the inherent uncertainties associated with climate change. Our goal is not only to assess the optimal investment strategy under average future weather conditions but also to evaluate the requirements for extreme weather scenarios. This approach enables a comprehensive understanding of how investment strategies might need to adapt to both typical and atypical climatic conditions. In this paper, we explore the implications of both demand and supply side shocks on energy system planning using a bottom-up capacity expansion model, focusing on the power system planning for Italy in 2030. 
%To our understanding, while existing literature has addressed the impacts of climate change on this nation, it remains limited and often lacks a comprehensive and thorough modelling framework \cite{tina_assessment_2021}. 
In this study, we build upon previous work by employing an integrated modelling approach to provide a more complete and rigorous analysis. We optimise the least-cost electricity system across various weather scenarios to account for potential future climatic conditions. By defining various weather scenarios for 2030, we can evaluate the impacts of both average near-future climate shifts and more extreme weather conditions. We combine existing and new empirical evidence to expand the understanding of power demand shocks and supply impairments under future climate change, with a focus on the implication of increased temperatures on air-conditioning demand, thermal power outages and hydropower generation. 
%While we also assess the impacts of climate change on solar and wind energy generation, these effects are not incorporated into the main capacity expansion run evaluated in this study ...
%, due to the lack of a definitive solution from empirical evaluations, a limitation previously noted in the literature \cite{bloomfield_quantifying_2021, russo_forecasting_2022,hu_implications_2023, tobin_climate_2016, jerez_impact_2015}. Nevertheless, for the sake of completeness, we provide an assessment of the impacts of average climatic shifts on solar and wind in a set of supplementary model runs, exposed in Appendix \ref{subsubsec: vre impacts}. 

The remaining sections of this paper are organised as follows. Section \ref{sec: materials and methods} outlines the methodology used, including the empirical evaluation of climate change impacts on power supply and demand, as well as the development of the bottom-up model for the Italian power system. Section \ref{sec: results} presents the findings of the study, while Section \ref{sec: discussion} provides a discussion of these results. Finally, Section \ref{sec: conclusions} outlines the possible implications for policy-making, and addresses the study's limitations along with recommendations for future research.

\medskip

\section{Materials and methods}
\label{sec: materials and methods}
\subsection{Italian power system}
\label{subsec: Italian power system}

In order to evaluate the effects of climate change on the Italian power sector in 2030 we adopt a power system modelling framework developed based on the open source model \textit{oemof}, which consistently represents electricity dispatch and optimisation (the model is described in details in Appendix \ref{appendixC}). A version of the power system model is available on \textit{GitHub} at \cite{Di_Bella_Oemof_Italy_2022} and was already employed in peer-reviewed studies \cite{di_bella_demand-side_2024, di_bella_power_2024}. The Italian power system is described with an hourly resolution, divided into the seven market zones defined by the transmission system operator Terna \cite{Terna7zones,Terna7zones_allegato}. The model includes the existing power generation capacities up until the year 2021, then the optimiser is able to install new power plants to minimise the total system costs within the limits of model constraints (outlined in Appendix \ref{appendixC}). In the majority of the outcomes, the optimisation is performed assuming the implementation of the mitigation policies for Italy, which are legally binding according to the European Climate Law \cite{noauthor_eur-lex_nodate}. The European decarbonization goal for 2030 is to decrease carbon dioxide released in the atmosphere by 55\%, in line with the \textit{Fit-for-55} policy package \cite{FitFor55}. A larger effort is expected from sectors in the EU Emission Trading Scheme (which are electricity and heat generation, energy-intensive industry sectors, aviation and maritime transport) reaching a 62\% decrease in CO$_2$ emissions \cite{noauthor_our_nodate,noauthor_what_nodate}. In particular, the power sector can count on more mature technologies than other energy sectors and it is a fundamental enabler of the transition of other energy sectors through electrification \cite{noauthor_secure_nodate}. Therefore, in this model we impose to the Italian power system in 2030 a reduction of CO$_2$ emissions of 65\% of electricity emissions with respect to their value in 1990 (they were 124.6 Mton of CO$_2$ \cite{ISPRAemissioni}). Additionally, in Section \ref{res: mitig vs non mitig}, we present scenarios in which no mitigation policies are applied (denoted as $Non Mitigated$ cases) to evaluate the implications of adapting the power system for climate resilience compared to the $Mitigated$ cases. 

The various elements of the power system necessary to shape the Italian electricity system are delineated in the Appendix A. Key features include: i) multiple market zones linked in the model through high-voltage transmission lines; ii) power production resources including natural gas, rooftop and utility scale photovoltaic, wind onshore and offshore, run-off-river, reservoir hydro, biomass and geothermal generation and imported electricity; iii) the possibility to add storage capacity for lithium-ion batteries and hydrogen storage technology, composed of electrolysers, fuel cells and hydrogen tanks. The model outcome has been validated with the data from the Transmission System Operator for the baseline year of 2019 \cite{noauthor_download_nodate}.  \\

\subsection{Climate variables}
\label{subsec: Climate scenarios}

\subsubsection{Data}

Historical weather patterns are derived from weather reanalysis data for the forty years from 1981 to 2020 (following a method from \cite{antonini_historical_2023}), available from ERA5 and ERA5-Land \cite{hersbach2020era5}. We  chose a time period of several decades centred around 2000 to strike a balance between - from the one side - evaluating weather conditions close to recent years and - from the other side - including several decades of weather data in order to derive reliable distributions of weather patterns. Nevertheless, we show that the impacts we estimate are largely unchanged if we restrict the time span of our historical weather scenario to the two decades of 2001-2020 (see Supplementary Information and Supplementary Figure \ref{Fig:base_year_comparison}). 
\\
The future climate projections derive from one CMIP5 EUROCORDEX Global and Regional Climate Model (GCM and RCM) (the GCM ICHEC-EC-EARTH and the RCM KNMI-RACMO22E). This data was preferred over CMIP6 climate projections because the former has hourly resolution, allowing us to conduct an assessment of climate change impacts retaining the high temporal frequency of the historical reanalysis ERA5 data. We consider as main future emission scenarios the RCP 4.5 (we find that no substantial differences arise with respect to hotter climate scenarios, e.g. RCP 8.5, in the time horizon of our analysis, centred around 2030, see Supplementary Figure \ref{Fig:rcp_comparison}). We match the historical period to the ERA5 reanalysis baseline period (1981-2020) and consider as future period the two decades around 2030 (2021-2040). In this way, our climate change impact projections should be considered as medium-term shifts in the climate occurring over three decades. While mean climate changes over only three decades might lead to negligible shifts in the mean climate conditions, the focus of this work is on both the mean and the tails of the weather distribution, allowing to test if power system planning will be impacted by climate change even in the next few decades, due to an amplification in extreme weather conditions. Both historical ERA5 reanalysis and future CORDEX projections data are taken from the population-weighted dataset of \cite{secures}, available for Italy at the NUT3 level.\\

\subsubsection{Definition of scenarios}

The impacts of weather patterns on the supply- and demand-side of the electricity system are included in the paper by considering four alternative scenarios
%. The first scenario (\textit{Historical Mean, $HM$}) is developed based on historical averages of weather patterns observed over four decades. An alternative scenario (\textit{Historical Extremes, $HE$}) takes into account historical extreme weather patterns, i.e. is developed to evaluate the power system implications of extreme weather events as they have occurred in the past. Both scenarios are developed as a reference for investigating the potential medium-term (around 2030) amplification of weather impacts on the power system when climate change is not accounted for. The third scenario (\textit{Future Mean, $HM$}) - which can be directly compared to the scenario $HM$ - considers only the climate change induced shift in the average weather patterns. Finally, the fourth scenario (\textit{Future Extremes, $FE$}) considers the climate change induced shifts in extreme weather events and can be directly compared to the scenario $HE$. These four scenarios - described in detail below - are 
developed separately for each impact category: demand for power, hydropower supply, thermal power supply and VREs power supply.

We start by considering a number $j$ of weather variables ($W^j$) - such as daily maximum temperatures -  at the NUTS3 level ($n$) for each hour ($h$), calendar day ($d$) and year ($y$). From each weather variable $W^j_{n,h,d,y}$ we derive the power system impacts based on different methodological approaches depending on the impact type (described in detail in Section \ref{subsection:demand_shocks} and Appendix \ref{subsec: Thermal generation outages}). Regardless of the specific approach, we can generalise the method adopted for the computation of impacts as follows. Consider the location- and time-specific impacts $\Upsilon_{n,h,d,y}$ computed from the weather variable $W^j$ through a damage function $f()$:

\begin{equation}
  \Upsilon_{n,h,d,y}  = f(W^j_{n,h,d,y})
\end{equation}    

First, we compute the impact of mean historical weather on the power generation variables (scenario \textit{Historical Mean, $HM$}) by taking the average over all years $y$ of $f(\hat{W}^j_{n,h,d,y})$ for a given location, hour and calendar day, where $\hat{W}^j_{n,h,d,y}$ is a weather variable of the ERA5 weather reanalysis spanning from 1981 to 2020:

\begin{equation}
  \Upsilon^{HM}_{n,h,d}  = \frac{\sum_{y}f(\hat{W}^j_{n,h,d,y}) }{y}
\end{equation}\label{eq:impact_hm}   

Second, we use the delta change-method, adopted extensively in the climate impact literature (see for instance \cite{mistry2017simulated,vavrus2015interpreting,arnell1996effects}) to calculate the impact due to the shift from the mean historical weather to the mean weather around 2030 (scenario \textit{Future Mean}, $FM$). More in detail, we compute the climate change amplification ($\Delta$) as the difference between the  future and historical EUROCORDEX projections of each weather variable $Z^j_{n,h,d,j}$ for $n$, $h$ and $d$, where means are computed over 20 years around 2010 $k$ (2001-2020)\footnote{We test two different historical periods baselines, alternatively 1981-2020 and 2001-2020. We find negligible differences in the amplification effect of climate change, as reported in Figure \ref{Fig:base_year_comparison}. } and 20 years around 2030 (2021-2040) $m$:

\begin{equation}
  \Delta^j_{n,h,d} = \frac{\sum_{m} Z_{n,h,d,j} }{m} - \frac{\sum_{k} (Z_{n,h,d,k}) }{k}
\end{equation}

To identify the impact of the scenario \textit{Future Mean}, we add the amplification $\Delta$ to $\hat{W}^j_{n,h,d,y}$ and, similarly to equation \ref{eq:impact_hm}, we take the average impact over the years $y$ (note that $\Delta$ is common across all years):

\begin{equation}
  \Upsilon^{FM}_{n,h,d}  = \frac{\sum_{y}f(\hat{W}^j_{n,h,d,y}+ \Delta^j_{n,h,d}) }{y}
\end{equation}\label{eq:impact_fm}

The implementation of $\Upsilon^{HM}_{n,h,d}$ and $\Upsilon^{FM}_{n,h,d}$ on a power system model allows to compare how slowly moving climatic changes in the mean weather conditions can affect the optimisation of power system capacity planning. Then, we evaluate the implications of weather-related impacts that occur in the tails of the historical and future simulated weather distributions, respectively. The two alternative extreme weather scenarios ($HE$ and $FE$) are constructed by considering the tail of the distribution of possible weather impacts resulting from $f(\hat{W}^j_{n,h,d,y})$, based on a quantile function $Q^{\chi}()$ and a quantile threshold $\chi$. We consider as alternative values for $\chi$ the 75th and 95th percentile of impacts. While the impacts in the model are identified at the regional and hourly level, the computation of the percentile is done by ranking each possible weather simulation at the annual and country level. In other words, we rank each of the possible weather realisations based on their aggregated country-level impact. We then select the 10 and 2 years which have the highest simulated impact for the 75th and 95th quantile thresholds respectively. Finally, we take the average the hour- and calendar-day specific impact in the subset of selected years, with the aim of reconstructing the simulated extreme weather conditions while at the same time maintaining a plausible inter-annual variability deriving directly from the observed reanalysis data characterising the selected years.

Note that when we evaluate each impact category in isolation we simulate shocks occurring once every four years (75th percentile) or once every twenty years (95th percentile). On the other hand, the model run that combines extreme weather shocks of all impact categories (demand, hydro-power and thermal generation) has a different implied probability given that some years in our distribution exhibit a pattern of co-occurrence of extreme weather impacts across multiple categories (as reported in Table \ref{tab: scenarios table}).\\

In the case of the \textit{Future Extreme}, the amplification of climate change is accounted for by adding the $\Delta$ amplification as in Eq. \ref{eq:impact_fm}. 

\begin{equation}
 \Upsilon^{HE,\chi}_{n,h,d}  = \frac{\sum_{p(y)^{\chi}}f(\hat{W}^j_{n,h,d,p(y)^{\chi}}) }{p(y)^{\chi}}  \end{equation}    

\begin{equation}
  \Upsilon^{FE,\chi}_{n,h,d}   = \frac{\sum_{z(y)^{\chi}}f(\hat{W}^j_{n,h,d,z(y)^{\chi}}+ \Delta^j_{n,h,d}) }{z(y)^{\chi}} 
\end{equation}    

Where $p(y)^{\chi}$ and  $z(y)^{\chi}$ are the subset of years among the historical and future simulations selected, respectively, depending on the given percentile threshold $\chi$.

%Figure \ref{fig:method_graph} presents a summary of features of the scenarios evaluated in this study:

%\begin{figure}
%    \centering
%    \includegraphics[width=1\linewidth]{figures/method_graph.png}
%    \caption{Enter Caption}
%    \label{fig:method_graph}
%\end{figure}

Table \ref{tab: scenarios table} presents a summary of the method used to compute the different weather impact scenarios:

\begin{table}[H]
    \centering
    \footnotesize
    \caption{Computation of alternative weather impact scenarios}
    \begin{tabular}{c c c c c }
    \hline
    \textbf{Acronym}&\textbf{Period}&\textbf{N. years}&\textbf{Statistic}&\textbf{Equation}\\
    \hline
     HM & Historical  & 40 & Mean of all years &  $\frac{\sum_{y}f(\hat{W}^j_{n,h,d,y}) }{y}$ \\
     HE & Historical  & 10 & Mean of years above 75th percentile & $\frac{\sum_{p^{75}}f(\hat{W}^j_{n,h,d,p^{75}}) }{p^{75}}$ \\
     HE & Historical  & 2 &  Mean of years above 95th percentile & $\frac{\sum_{p^{95}}f(\hat{W}^j_{n,h,d,p^{95}}) }{p^{95}}$  \\
     FM & Future  & 40 & Mean of all years & $\frac{\sum_{y}f(\hat{W}^j_{n,h,d,y}+ \Delta^j_{n,h,d}) }{y}$ \\
     FE & Future  & 10 &  Mean of years above 75th percentile & $\frac{\sum_{z^{75}}f(\hat{W}^j_{n,h,d,z^{75}}+ \Delta^j_{n,h,d}) }{z^{75}}$ \\
     FE & Future  & 2 &  Mean of years above 95th percentile & $\frac{\sum_{z^{95}}f(\hat{W}^j_{n,h,d,z^{75}}+ \Delta^j_{n,h,d}) }{z^{95}}$ \\
    \hline
    \end{tabular}
    \label{tab: scenarios table}
\end{table}

\subsection{Impacts on electricity demand}
\label{subsection:demand_shocks}

The response of the hourly electric load to temperature is computed taking into account two aspects: the short-run co-variation between load and weather as well as the long-run amplification in the short-run load-weather relationship due to the growth in the adoption of air-conditioning appliances in the residential sector. First, in order to identify the increase in electricity demand related to the growth in AC uptake across Italian households we use an empirically estimated response function that provides a generalised functional relationship between daily maximum temperatures, AC market saturation and peak load demand, provided by \cite{colelli2023}. We project the prevalence of residential AC ownership rates in Italy at the regional (NUTS 2) level by associating for each location and year the probability of AC ownership in households based on the adoption function of \cite{colelli2023} (the detailed equation used by this study are shown in the Supplementary Information). 

Historical AC ownership rates at the regional level are taken from the Italian Budget Survey published by ISTAT for the year 2019 \cite{ISTAT_HBS}. The projected regional AC share is shown in the Supplementary Figure \ref{Fig:regional_ac_share}. In general, the average AC ownership rate in Italy goes from 33\% in 2019 to 47\%-55\% in 2030 under RCP 4.5-8.5.  The regions with the highest current as well as future AC share are the ones characterised by higher annual CDDs (Sardegna, Sicilia, Puglia) or by higher income per capita levels (Veneto, Emilia-Romagna, Lombardia). 

Second, we couple the future NUTS3-level hourly temperature projections provided by \cite{secures} and the NUTS2 AC prevalence levels to estimate the amplification of hourly electricity demand due to climate change, based on the temperature-load coefficients ($\widehat{\beta}$) resulting in the non-linear function $h()$ estimated in \cite{colelli2023} (taking the form shown in the Supplementary Methods), and the projected hourly temperatures provided by \cite{secures}. As in \cite{colelli2022impacts} the non-linear temperature-load function $h()$ is defined based on the population-weighted temperatures binned into $k$ intervals of 3$^\circ$C width, $B_k=[\underline{T}_k, \overline{T}_k)$, here we construct a $k$-vector of indicators that track whether each hour's mean temperature falls within a given interval:
$$T^k = 1 \cdot \lbrace T \in B_k \rbrace + 0 \cdot\lbrace\text{Otherwise}\rbrace$$. 

The observed historical binned hourly temperature for each location ($n$) for each hour ($h$), calendar day ($d$) and year ($y$), $\hat{T^{k}_{n,h,d,y}}$ and the difference in the EUROCORDEX climate series for the historical and future epochs $\Delta(T^{k}_{n,h,d})$ are used to project the impact on the hourly load based on the historical and future AC prevalence levels, respectively, in each scenario, based on the equations presented in \ref{subsec: Climate scenarios}.

\begin{figure}[H]
\begin{center}
\includegraphics[scale=0.6]{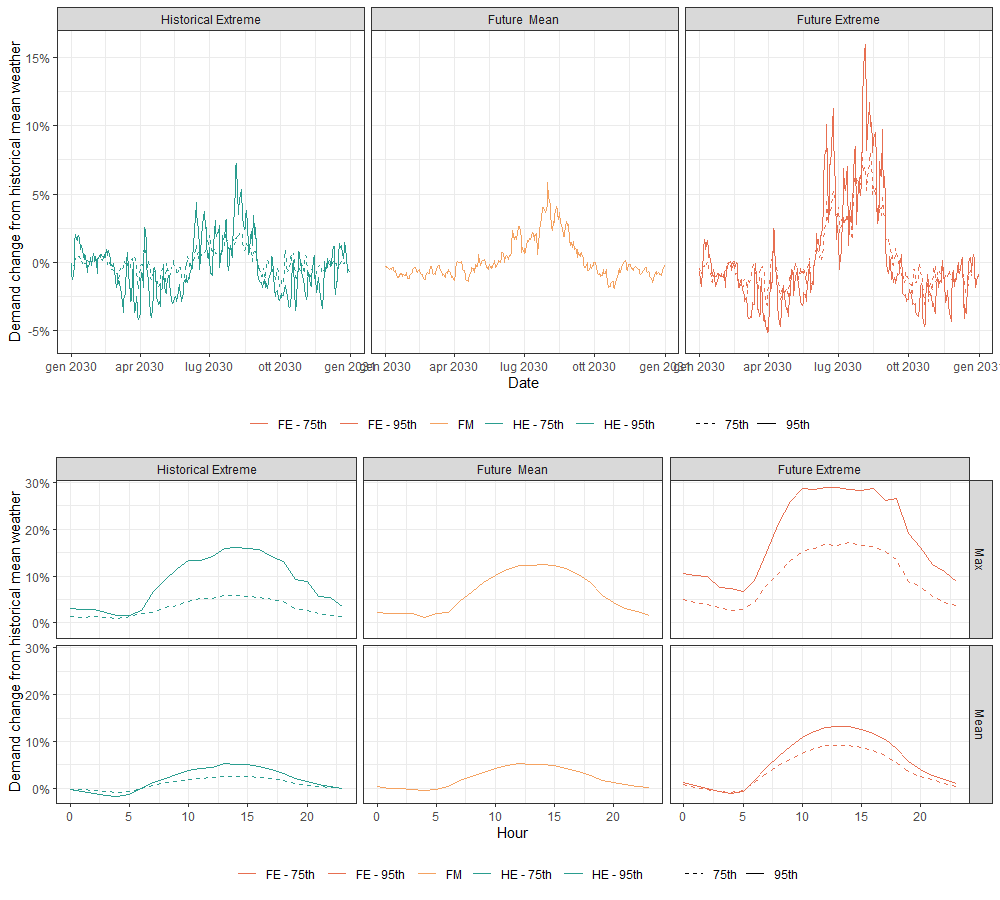}
\caption{Climate change impacts on electricity demand. Panel above: Changes in daily load with respect to the mean weather scenario, in each alternative climate scenario. 
Panel below: Changes in hourly load n the summer with respect to the mean weather scenario, in each alternative climate scenario. Solid lines show the change of the 95th percentile, while dashed lines of the 75th percentile.}
\label{Fig:demand_impacts}
\end{center}
\end{figure}

Figure \ref{Fig:demand_impacts} Panel a shows how the mean temperatures vary in each season depending on the scenario. A large increase is observable in mean temperatures during summer but also a higher frequency of warm temperatures in winter. Panel b shows the resulting amplification of the hourly load in the summer around 2030, with respect to the \textit{Historical Mean} scenario, averaged for all bidding zone. Maximum and mean amplifications of the hourly load at the national level reach over 30\% and 20\% in the central hours of summer months, respectively.
\begin{comment}

An important note has to be done on the impacts of weather variable on electricity demand for heating. This is visible in Figure \ref{Fig:demand_impacts} panel a for the winter months. The correlation for the effects of temperature on power load for heating are produced using the whole Europe as a sample, thus they include varius levels of heating electrification. In Italy, renewable energy sources cover around 20\% of the consumption for heat, with a target in 2030 of 33.1\% \cite{PNIEC, franci_ruolo_nodate}. Currently the final consumption for space heating met with electricity is 1 TWh \cite{noauthor_statistics_nodate}; projections assume that the number and capacity of heat pumps in Italy will double by 2030 \cite{croci_il_nodate}, we can assume to also double the consumption, thus 2 TWh would be around 2\% of the winter consumption of electricity.
\end{comment}

\subsection{Impacts on power generation}\label{subsection:supply_shocks}

In this work we take into account the potential impact of long-run climate change and short-run weather variability on power generation in several dimensions. 

\begin{figure}[H]
\begin{center}
\includegraphics[scale=0.5]{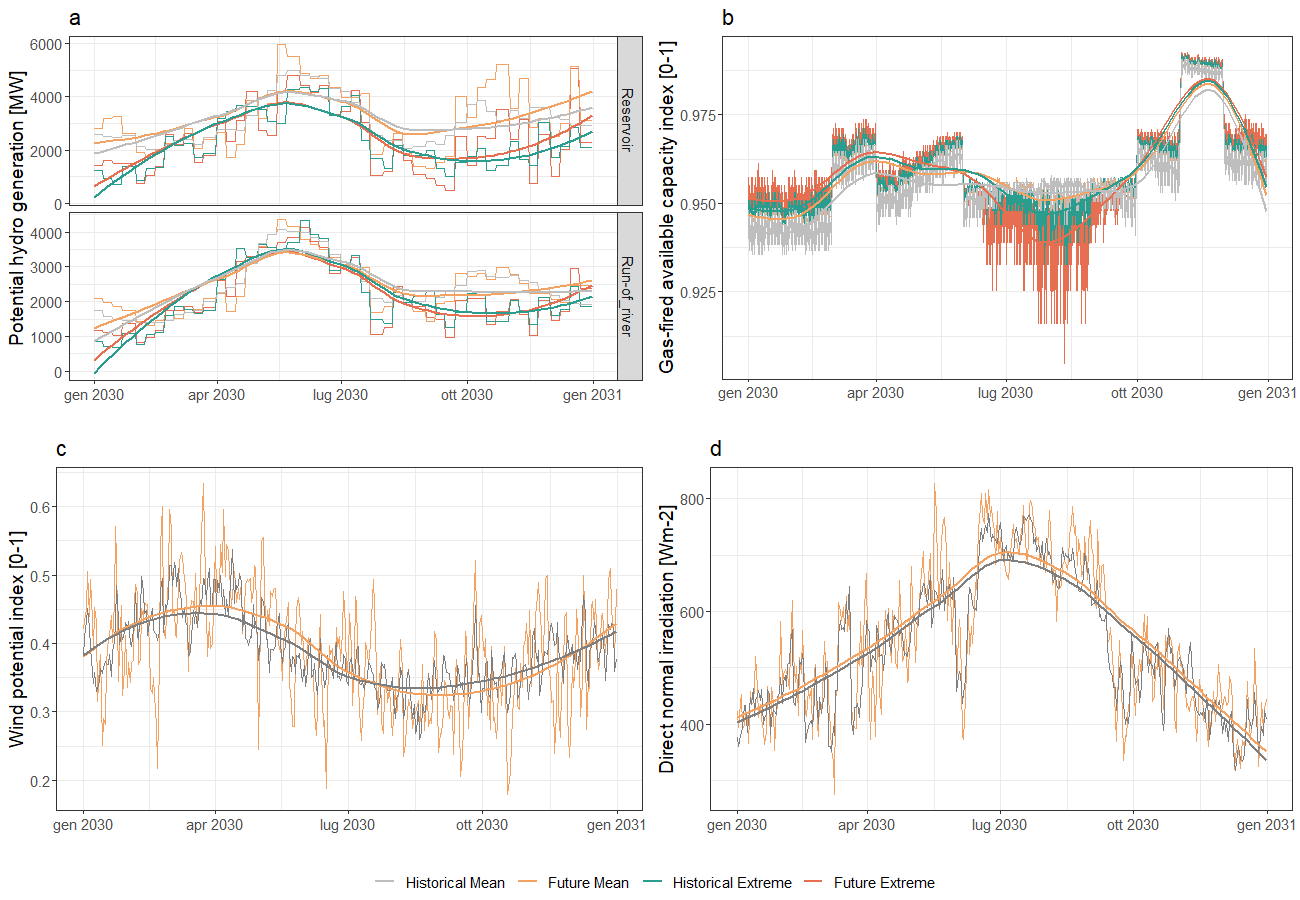}
\caption{Climate change impacts on power generation. Panel a: national-level potential hydro-power generation across the four climate scenarios; Panel b: Normalised availability of gas-fired generation after accounting for the occurrence of unplanned outages, across the four climate scenarios; Panel c: Normalised wind power potential at the national level, in the two selected scenarios; Panel d: BNI averaged at the national level, in the two selected scenarios. Impacts shown in Panel a, c and d are computed using as input data \cite{secures}. Impacts shown in Panel b are developed based on the econometric analysis presented in Appendix \ref{subsubsec: empirical analysis}. The values used as input data to the model are shown as thin lines, while thick lines show the smoothed time series for enhancing the inspection of trends in the data.}
\label{Fig:generation_impacts}
\end{center}
\end{figure}

First, we develop a set of projections on the changes in hydro-power generation by exploiting the daily time series of hydro-power potential at the NUTS3 level developed in \cite{secures}, distinguishing between run-off-river and reservoir. The potential hydro-power generation is combined with the installed production capacity of hydro-power plants. We emphasise that this analysis does not incorporate changes in the management of water demand for human needs, as they fall outside the study's scope. However, it is worth noting that such changes could potentially mitigate the impact of decreased water availability on hydro-power production \cite{giuliani_coupled_2016,herman_climate_2020}. We compute both daily and weekly power generation levels in each scenario (\textit{Historical Mean}, \textit{Future Mean}, \textit{Historical Extreme}, \textit{Future Extreme}) for each Italian macro-zone (see Figure \ref{Fig:generation_impacts} Panel a for the projected value of the country-level total hydropower generation).

Second, we develop a statistical analysis to investigate the impacts of daily maximum temperatures and water runoff anomalies on the availability of thermal power generation. More in detail, we generate a regression model based on unexpected outages data collected from 2018 to 2022 in Italy. The dataset \cite{entsoe_outages} includes information of over 4000 outages, of which 2352 of gas- and 1887 of coal-fired generation units. The method adopted partially follows a previous analysis (\cite{coffel2021thermal}), but expands from the literature by providing country-specific and fuel-specific responses. In the most detailed specification, we identify fuel-specific impacts by including a set of interaction terms between temperature and runoff anomalies and a character variable representing the power plant type. Comprehensive information on the methods employed is provided in Appendix \ref{appendixB}. We find that the likelihood of occurrence of an outage for coal-fired generation increases considerably when daily maximum temperatures surpass 35°C, reaching 10-50\% at 40°C, depending on the water runoff anomaly. Gas-fired generation is less sensitive to high temperature and low water runoff levels, but the likelihood of an outage is non-negligible and between 5-25\% at 40°C.
Since in our power model optimisations coal generation is phased out from the mix, we focus on the projections of impacts for gas-fired generation. We use the estimated outage occurrence function in conjunction with daily maximum temperature anomalies at the location of the gas-power plants to simulate daily thermal outage occurrence. Finally, we aggregate the resulting projections of power plant outages at the bidding-zone level to obtain an indicator of power generation availability changes around 2030 in the four scenarios identified in section \ref{subsec: Climate scenarios} (see Figure \ref{Fig:generation_impacts} Panel b).

Our model has a deterministic approach in the quantification of the available supply of power generation sources. As noted by \cite{oree2017generation}, the main problems of this method relate to its unique economic objective function and the absence of uncertainties in its formulation. Other works (e.g. \cite{jin2014temporal}) have proposed a stochastic model where long-term uncertainty in the VREs resources is introduced by using multiple scenarios consisting of weekly time series of hourly wind power output data. We leave the adoption of a stochastic approach to future studies, and we partially address this limitation by observing the outcomes of two alternative set of model runs: one with no impact of climate change on VREs and one where the generation profiles of VREs in the four different scenarios are included  - although through the deterministic approach. We observe negligible differences when incorporating the effects of meteorological variables on VREs (see Appendix \ref{subsubsec: vre impacts}). However, as this approach is not ideal for capturing VREs' uncertainty in capacity expansion analyses, we have chosen to present our main findings without these considerations.

\section{Results}
\label{sec: results}
\subsection{Installed capacity}

\begin{figure}[H]
\begin{center}
\includegraphics[scale=0.7]{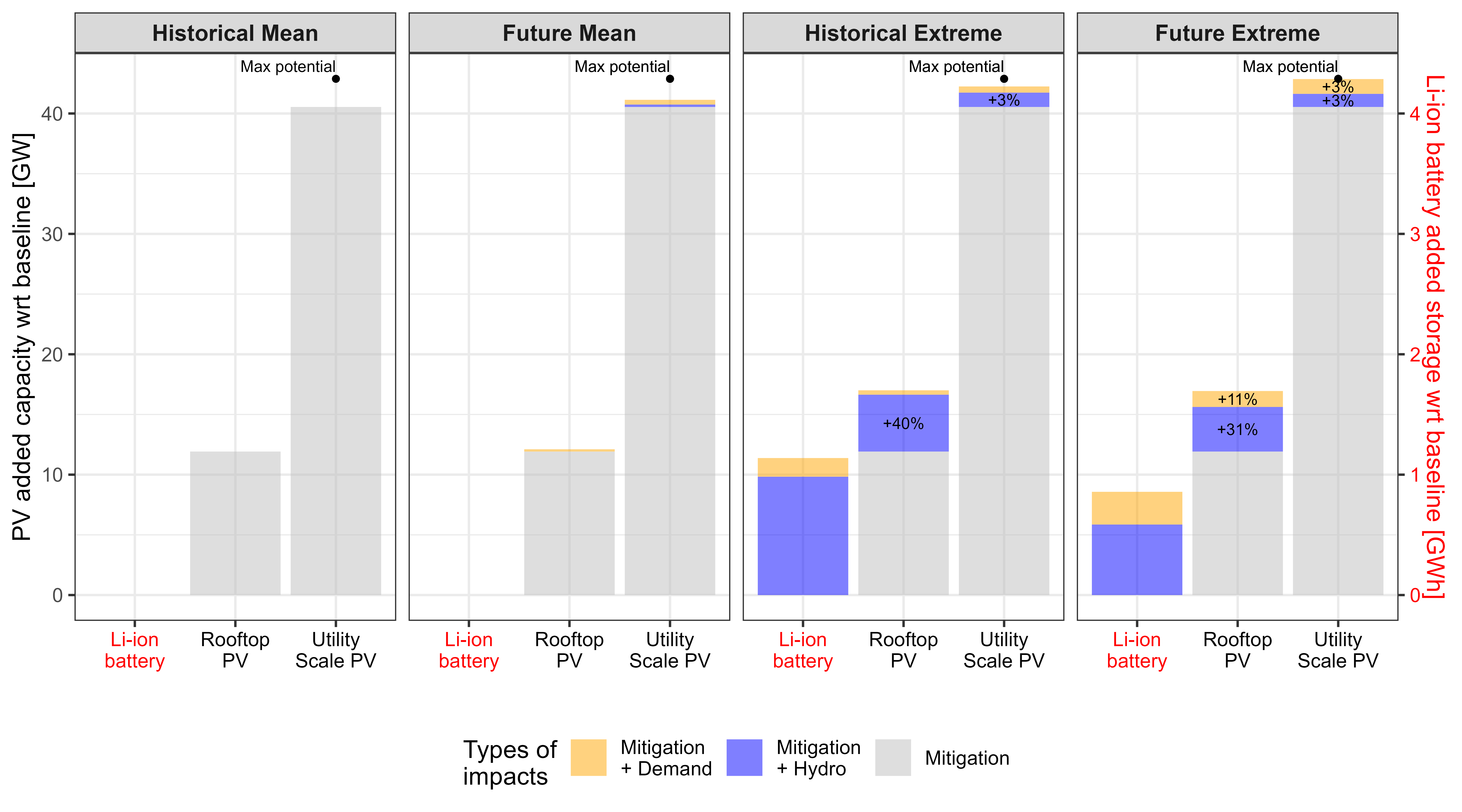}
\caption{Added capacity for PV power (utility scale and rooftop) and for lithium-ion storage with respect to the 2019 Italian power system baseline (described in Table \ref{tab: renewable capacity 2021} for the four scenarios. The different colours highlight the climate driver behind each capacity expansions. Grey bars show the added capacity when the only effect is the push to reach decarbonization goals, while in the others each weather variable is evaluated on top of the mitigation. In addition to the impact of mitigation measures, the purple bars encompass the marginal effect of weather variables on hydroelectric generation, while the yellow bars on power demand.}\label{Fig:added_cap}
\end{center}
\end{figure}

We identify considerable impacts of mitigation policy goals and climate impacts on the optimal generation capacity. The installed capacities for the 2030 Italian power system, determined through investment optimisation, meet hourly load requirements while enforcing a 65\% reduction in power sector emissions. We consider as "added capacity" the additional one optimised by the solver on top of the existing one in the 2019 reference case, detailed in Table \ref{tab: renewable capacity 2021}. The effects of mitigation (grey bars) and of climate change and extreme weather (coloured bars) on generation and storage capacity are displayed in Figure \ref{Fig:added_cap}, for the different climate scenarios and generation technology.

In Figure \ref{Fig:added_cap}, the panel on the left shows the changes in installed capacity resulting from the \textit{Historical Mean} scenario: in this case, the added capacity is driven only by mitigation policies. Around 50 GW of photovoltaic panels, mostly in large scale fields, are required to reach 2030 decarbonization targets for power generation in Italy in a cost optimal system. The alternative climate scenarios take into account the effects of long-term climatic changes and extreme weather variability on electricity demand, thermal power plants and hydro-power. We performed various optimisation accounting for these effects separately to grasp which has more consequences on the installed capacities. Despite increasing the likelihood of unplanned outages, the impact of climate change on gas-fired plants does not affect optimal installed capacity, thanks to the ample unused natural gas power capacity available in both the 2019 reference and mitigation scenarios \footnote{Note that such results depends on the modelling framework, which assumes perfect foresight of the reduction in available capacity from unplanned power outages. More detailed assessment of the impacts on short-term market operations when unplanned outages occur fall out of the scope of this analysis.}. As a side note, changes in the availability of solar and wind energy from the \textit{Historical Mean} to the \textit{Future Mean} climate scenarios are negligible when aggregating across time and space, and point to a modest increase in rooftop solar generation and curtailment to account for an increase in seasonal weather variability of wind generation (results are presented in the Appendix \label{subsubsec: vre impacts}).
In the \textit{Future Mean} scenario, considering only the effects of climate change from a shift in the mean climatic conditions, the additional installed capacity is minimal with respect to the \textit{Historical} scenario. This indicates that the photovoltaic power installed for mitigation purposes would be sufficient to ensure the system's resilience to an average year of weather under climate change conditions occurring by 2030. In fact, solar electricity is significantly correlated with power demand driven by AC appliances, since they are both relevant in the middle of the day (the contribution of solar PV on hourly power generation during summer months is shown more in detail in the next section, particularly in Figure \ref{Fig: hourly weeks net PV}). When considering historical extreme weather (\textit{Historical Extreme} panel), changes in the availability of hydropower have a relevant role in increasing the need for solar capacity (plus $\approx$ 5GW). Solar electricity is produced during the day, substituting the reduced run-off-river generation and shifting the use of reservoir hydro during night-time, instead of employing natural gas-fired plants. This mechanism allows the system also to stay compliant with the emission reduction target by reducing the use of fossil-based electricity.
Furthermore, climate-induced changes in hydro-power availability have remarkable effects in driving the uptake of short term storage. Indeed, when the availability of hydro power generation is reduced, the need for flexibility in the system is satisfied with li-ion batteries. In this \textit{Historical Extreme} case the storage capacity is relatively small ($\approx$ 1GWh), but for higher degrees of decarbonization a larger volume will be needed. To provide a comparison, Terna, as published in the 2022 Future Energy Scenarios document, foresees approximately 71 GWh of new utility-scale storage capacity will need to be developed by 2030 to meet the requirements of the \textit{Fit-for-55} scenario \footnote{The optimisation does not install extra storage technologies in the \textit{Future Mean} case since it works in perfect foresight and thus acts to minimize the costs, but it is not able to foresee market dynamics and future decarbonization of the system.}. Finally, the \textit{Future Extreme} scenario presents a stronger impact of the increase in demand for AC utilization. Utility Scale PV installation reach the maximum potential, taken from Trondle et al. \cite{trondle_home-made_2019}, which considers social and technical constraints to fields installations. The optimisation algorithm tends to prioritise Utility Scale panels over rooftop PV, since they are more costs efficient due to their lower price for economies of scale, but land availability, protected areas and social constraints limit strongly their potential. A sensitivity analysis on the value for its maximum potential might enlarge the use of this technological option, given its price competitiveness. In the \textit{Future Extreme} scenario, the effect of reduced hydro generation on installed capacity is lower than in the \textit{Historical Extreme}  scenario because of the balancing between summer and winter water inflows, observed in the mean climate change weather analysis (Figure \ref{Fig:generation_impacts}). As the impact on hydro electricity production goes down, also the necessity for substitute li-ion batteries decreases (half of the need in the \textit{Historical Extreme} case driven by hydro power).

\begin{table}[H]
    \centering
    \footnotesize
    \caption{Investments related to the additional installed capacities for generation and storage technologies in the four weather scenarios [M€]. In brackets we show the percentage change with respect to the investments in the \textit{Historical Mean} scenario.}
    \begin{tabular}{c c c c c }
    \hline
    \textbf{Technology}&\textbf{Historical Mean}&\textbf{Future Mean}&\textbf{Historical Extreme}&\textbf{Future Extreme}\\
    \hline
     Utility Scale PV & 2312.7 & \makecell{2346.6 \\(+1.5\%)} & \makecell{2414.9 \\ (+4.4\%)} & \makecell{2455.1 \\ (+6.2\%)} \\
     \hline
     Rooftop PV & 1115.6  & \makecell{1135.8 \\ (+1.8\%)} & \makecell{1595.5 \\ (+43.0\%)} & \makecell{1589.9 \\ (+42.5\%)} \\
     \hline
     Li-ion battery & 0.0 & 0.0 & 16.5 & 11.6 \\
    \hline
    \end{tabular}
    \label{tab: investments table}
\end{table}

In Table \ref{tab: investments table} we display the values of annualised investments related to the extra installed capacities represented in Figure \ref{Fig:added_cap}. These values reflect the compound effect of all the weather variables in the optimisation (AC demand, hydro and thermal power). %unlike in Figure \ref{Fig:added_cap} where we can see the decomposed effect based on types of impacts. 
The investment values reflect the results already observed in the figure: in the \textit{Future Mean} scenario, there is a negligible increase in costs compared to the \textit{Historical Mean}. Given that all scenarios assume a 65\% reduction in CO$_2$ emissions from power generation, this indicates that decarbonising the energy system may yield significant synergies with adaptation strategies to climate change. This is largely attributed to the integration of climate-resilient technologies, particularly rooftop PV systems and utility-scale PV installations. In the extreme weather cases (\textit{Historical} and \textit{Future}), the primary increase in system expenses is attributable to rooftop PV installations. These technologies are more costly compared to Utility-Scale solar panel fields; however, rooftop PV installations are essential due to the larger availability of potential spaces for their installations and their ability to reduce centralised load demands. Indeed, rooftop PV is a form of distributed energy production and plays a crucial role in mitigating stress on the high-voltage grid by generating electricity locally. This becomes particularly advantageous under increased electricity demand scenarios due to climate change or extreme weather conditions, as it diminishes the dependency on centralised power sources, as demonstrated in Figure \ref{Fig: net demand after PV} in Appendix \ref{appendix: rooftop PV}.

\subsection{Electricity generation}

\begin{figure}[H]
\begin{center}
\includegraphics[scale=0.7]{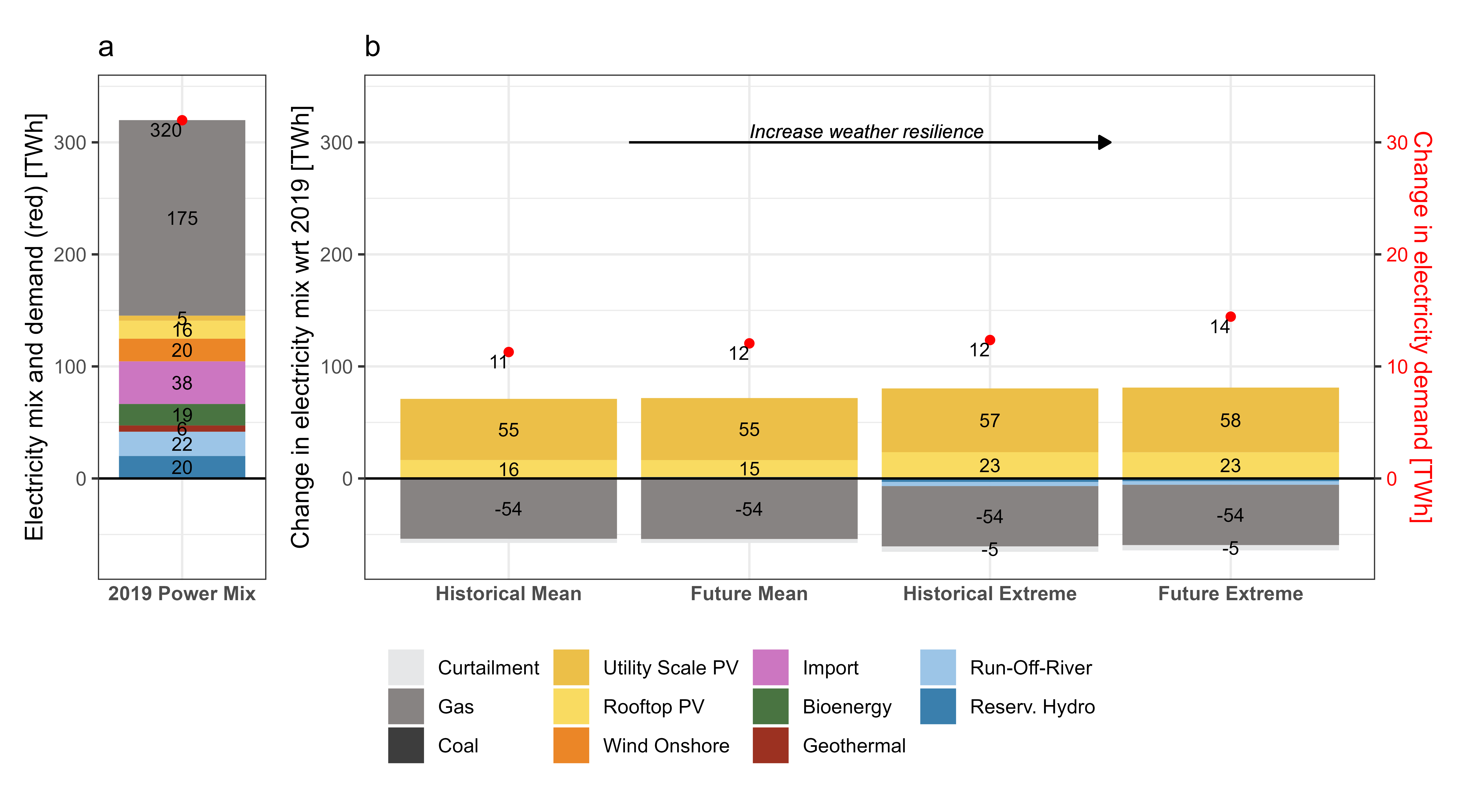}
\caption{Panel a: Model output of the electricity mix in Italy for the year 2019, different colour for each power technology. The red dot indicates the annual load in TWh; Panel b: Variation in the Italian electricity generation for year 2030 in the presented weather scenarios. Going from left to right, each output provides a stronger resilience to extreme weather conditions. In this panel the red dot precises the variation of national annual power load with respect to the 2019 baseline.}
\label{Fig:power_mix}
\end{center}
\end{figure}

As a consequence of the changes in the power capacity mix induced by mitigation policies\footnote{As a reminder, all the expansion capacity optimisations are performed with the underlying assumption of coal phase-out \cite{PNIEC} and an abatement of CO$_2$ emissions in line with European policies \cite{noauthor_fit_nodate}.}, the generation mix is substantially different in 2030 with respect to the reference generation of 2019 \footnote{The model outcome for the 2019 reference year has been validated with the data from the Transmission System Operator \cite{noauthor_download_nodate}.} (panel a of Figure \ref{Fig:power_mix}). 

Despite the cap on carbon dioxide emissions, natural gas still plays a significant role in 2030 power mix: 54 TWh less mean that 121 TWh of electricity are still gas-generated, one third of the total annual value. Solar electricity production has a relevant increase in all the scenarios. In the \textit{Historical Mean} case this is due to the abatement target for CO$_2$ that forces the model to install and use more solar panels. Instead, the marginal increase in solar power generation in the other weather scenarios is driven by the effect of climate change and more frequent severe weather events. Another crucial point is curtailment: in the case of extreme weather in both historical and future climate, excess electricity would be 5 TWh. Although curtailed power constitutes only about 1.5\% of the total demand, it requires attention as it presents both challenges and opportunities. This surplus cheap energy can be used within hard-to-abate sectors, reducing overall costs for mitigation solutions. On the other hand, excess power could be accounted for by load shifts, moving the request of electricity for certain services or appliances to hours of peak generation.\\

The total annual electricity demand (shown with a red dot in Figure \ref{Fig:power_mix}) grows from 320 TWh in 2019 to 331-334 depending on the climate scenario. In the \textit{Historical Mean} case, the additional electrical load is based on exogenous projections for the year 2030 \cite{TernaAdeguatezza}, which consider the electrification of end-uses, particularly for passenger transport, industry and heating. The projections of \cite{TernaAdeguatezza} also take into account the simultaneous reduction driven by investments in energy efficiency. The change in demand with respect to the \textit{Historical Mean} case is driven by the change in hourly demand for heating and cooling devices projected based on the method presented in section \ref{subsection:demand_shocks}. Changes in annual aggregate demand mask considerably higher impacts occurring at the hourly level in the summer.

\begin{figure}[H]
\begin{center}
\includegraphics[scale=0.5]{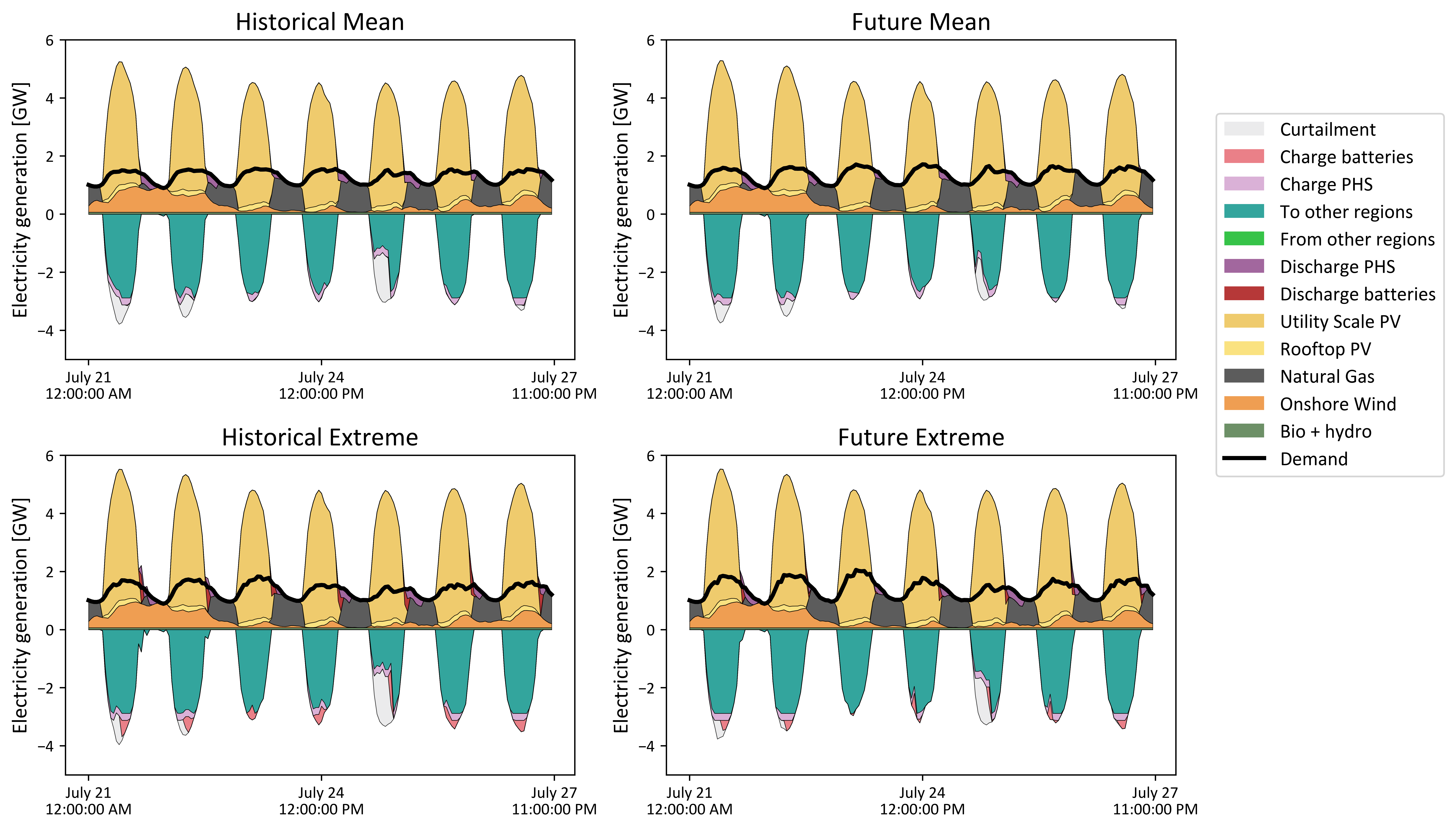}
\caption{Hourly generation and demand for Sardinia region during a summer week, considering the four different weather cases. The generation from Biomass, Run Off River and Reservoir Hydro is aggregated, as well as the one from and to other regions. Sardinia as a market zone in the Italian power system is connected to Centre-North, Centre-South and Sicily.}\label{Fig: hourly weeks net PV}
\end{center}
\end{figure}

Figure \ref{Fig: hourly weeks net PV} shows an example of hourly generation and demand in the four weather scenarios for a representative summer week (going from Monday to Sunday) in Sardinia. We chose this power market zone since, in the two extreme weather scenarios, it is where the majority of li-ion battery capacity is installed to take advantage of the excess renewable generation (the impact of solar PV on the net load of all the Italian regions during the summer is shown in Supplementarry Figure \ref{Fig:solar_net_load}). Inspecting the hourly behaviour of demand and production in each weather scenario confirms the capability of PV panels to produce a large amount of electricity during the central hours of the day, precisely when the demand for cooling is projected to peak. Additionally, it is noteworthy to notice the interactions with neighbouring regions (\textit{To other regions} and \textit{From other regions} in the legend, representing Centre-North, Centre-South and Sicily) and the utilisation of storage mechanisms, specifically the charging and discharging processes of PHS and batteries. This strategy involves charging the storage system with excess solar generation during midday, followed by discharging to supply power in the evening, reducing the use of natural gas.

\subsection{Mitigation vs adaptation: trade-offs and synergies}
\label{res: mitig vs non mitig}

While the main objective of this work is to investigate climate change impacts in a power system which is compliant with the legally binding European mitigation laws by 2030, in this section we compare those results with an alternative case in which climate change affects the Italian power sector when no stringent mitigation target is enforced. We undertake this analysis to determine whether achieving mitigation goals entails a trade-off with adaptation goals, meaning planning a system resilient to climate change and extreme weather conditions. To this aim, we run each climate impact scenario (\textit{Historical Mean}, \textit{Future Mean}, \textit{Historical Extreme} and \textit{Future Extreme}) assuming that the Italian power system faces no mitigation target in 2030 ($Non Mitigated$ scenario). The results in terms of costs are outlined in Figure \ref{Fig: cost diff mit no mit}, showing the total annual system cost in 2030 for a $Non Mitigated$ power system in the column on the left, then that cost for a $Mitigated$ system in the right column. The waterfall bars represent the cost differences due to the achievement of the mitigation targets, always including the adaption to the impacts of the specific weather scenario.

\begin{figure}[H]
\begin{center}
\includegraphics[scale=0.65]{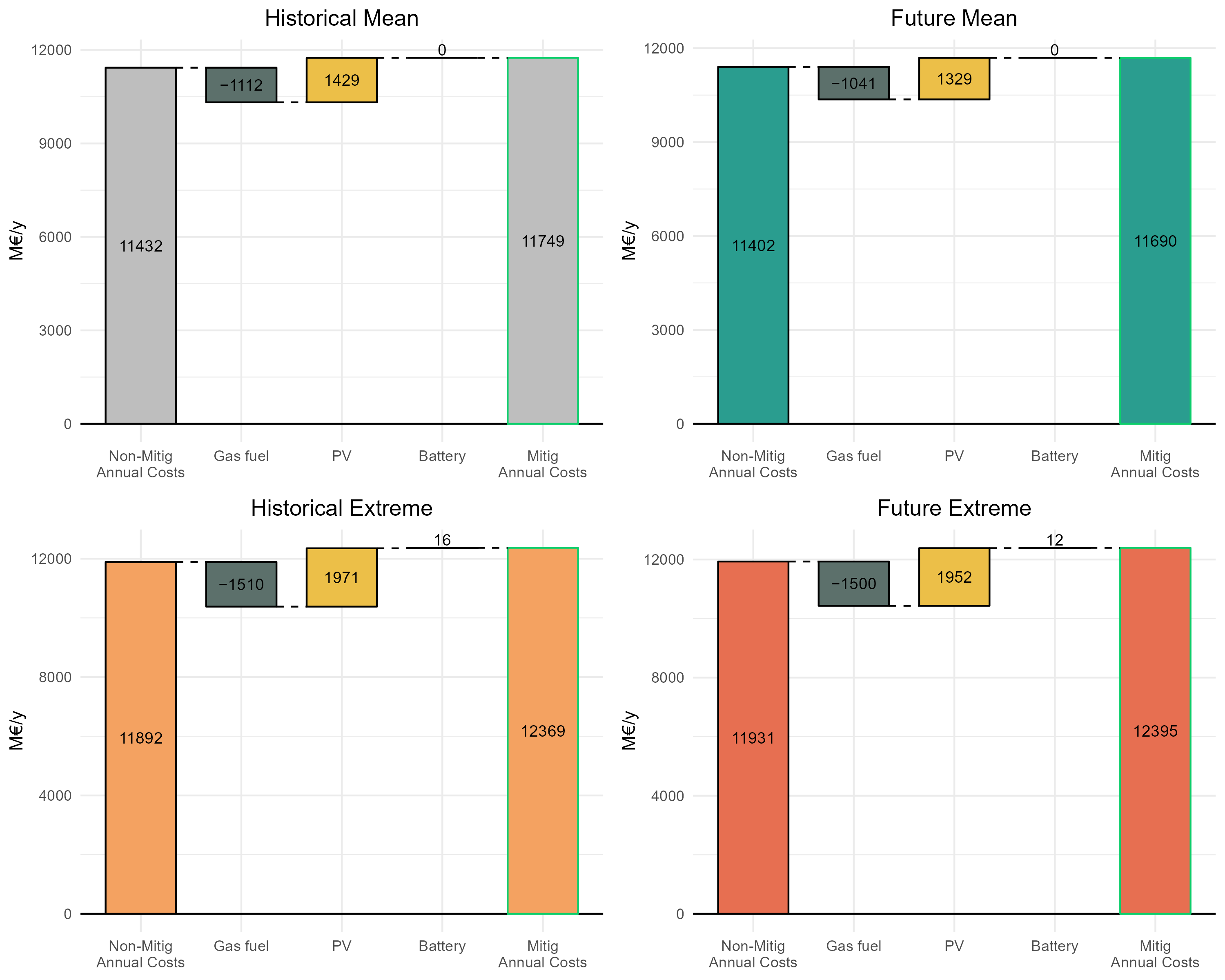}
\caption{Waterfall charts going from the total annual system cost for the $Non Mitigated$ case to the $Mitigated$ case (with green borders) in the four different weather scenarios analysed in the paper. The values on the bar between the one on the left and the on on the right in each graph describe the costs items differences to achieve a mitigated Italian power system starting from the $Non Mitigated$ but resilient one.}\label{Fig: waterfall mit no mit}
\end{center}
\end{figure}

As expected, the cost to purchase natural gas is always negative; instead, installation costs for solar technologies and batteries increase, even if the increase of the latter is negligible. The crucial difference is that the expenses for the acquisition of fuels is an operational cost, while the others are initial investment costs. This means that installing renewable technologies has a higher initial expenditure but significantly enhances the energy independence of the country and its resilience to geopolitical tensions and volatility of prices. In both the $Non Mitigated$ and the $Mitigated$ scenarios, total annual system costs increase when going to Extreme weather cases (upper vs lower graphs). When moving to the scenario resilent to the effects of climate change  (upper left vs upper right graphs), the expenditures decrease in the $Mean weather$ cases. The key message is that adapting to changing average weather patterns, particularly with regard to air conditioning demand and the unavailability of thermal and hydro power generation, reduces mitigation costs. Solar panels play a crucial role in this trade-off, as they not only represent an environmentally sustainable technology but also enhance the system's resilience to increasing temperatures. When considering extreme weather conditions, the effects of climate change result in a rise in expenditures (lower left vs lower right graphs). A pivotal aspect is that extreme weather events, which will become more frequent even under the RCP 4.5 scenario around 2030, pose a relevant threat and a possible increase in expenses for the power system. Importantly, the differences between total annual system costs in the $Non Mitigated$ and $Mitigated$ cases are quite negligible, highlighting the advantages of planning for a system concurrently decarbonized and adapted to a changed climate.

\section{Discussion}
\label{sec: discussion}

In this study, we employ a capacity expansion model of the Italian power system projected for 2030 to assess the necessary investments in power generation capacity quantitatively. This analysis is conducted with the dual objective of meeting mitigation targets and addressing the impacts of climate change on both power demand and generation while maintaining a high temporal resolution. The European decarbonization goal for 2030 is outlined in the \textit{Fit-for-55} policy package and mandates a 55\% reduction in CO$_2$ emissions \cite{FitFor55}. The power sector, leveraging its advanced technologies, is expected to achieve more considerable emission reductions. Thus we assumed a 65\% decrease in Italian electricity CO$_2$ emissions compared to 1990 levels, capping them at 44 Mton of CO$_2$ in the year 2030. We develop four alternative meteorological scenarios to thoroughly decompose the effects of climate change, distinguishing between shifts in the mean and the extremes of the weather distribution while assuming the implementation of decarbonization policies.

Overall, we find that transitioning towards a low-carbon power system capable of meeting demand under extreme weather patterns -  both as occurred in the past and amplified by climate change - necessitates a substantial increase in installed capacity, specifically in solar generation and short-term storage. Around 5-8 GW of additional PV capacity and 0.8-1.1 GWh of li-ion batteries are required at increasing weather stress on the 2030 Italian power system. These expansion requirements serve to withstand, at the same time, higher future AC-induced peaks in the load as well as lower hydropower generation. Given the pivotal role that electricity is expected to play in driving the decarbonization efforts of hard-to-abate energy sectors, it becomes imperative to prioritise the ability of the power sector to address such concerns. While demand-side measures related to electricity and water use in agriculture and other end-use sectors could help alleviate stress on the electrical grid, they are outside the focus of this research.

A valuable feature of this work is that it examines the adequacy of the power sector while at the same time ensuring that decarbonization policies are implemented, therefore allowing to identify of potential trade-offs or co-benefits between adaptation and mitigation. The outcomes of the optimisations find that the largest installation of new generation technologies is of Utility Scale PV (40-42 GW) since this is the most cost efficient in the Italian peninsula (see Figure \ref{Fig:added_cap}). Rooftop PV requires an addition of 12 GW to meet decarbonization goals, and extra 5 GWs are needed in the most extreme weather cases. This decentralised technology is crucial in reducing the stress on the electric grid and satisfying the power load locally. Furthermore, we find robust evidence that accounting for extreme weather events (both in the \textit{Historical} and \textit{Future Extreme} scenarios) affects the optimal capacity mix: a power system resilient to extreme events requires storage to help balance the water scarcity and thus hydropower unavailability. Climate change and extreme weather might hinder the availability of programmable reservoir hydropower plants and non-dispatchable RES might not be sufficient to meet the hourly load. Investments in storage capacity could be fundamental to alleviate the burden on the system and to mitigate the volatility and unpredictability in power generation. In this study, two storage options are available for the power system model: a short term storage with li-ion batteries and a long-term one with hydrogen tanks, electrolysers and fuel cells. Given the short-term nature of meteorological impacts, lithium-ion batteries are preferred over hydrogen storage for energy management. Chemical batteries, such as lithium-ion, are particularly well-suited for daily and short-term energy storage due to their minimal energy losses during charge and discharge cycles \cite{ibrahim_energy_2008}, making them an effective alternative to weather-dependent hydro power. Moreover, installing electrolysers, pressurised tanks and fuel cells has a high cost, which becomes viable only in more severe abatement scenarios.

The generation mix is markedly different from the reference generation of 2019, with solar electricity production showing substantial increases across all scenarios. This rise is attributed to the stringent CO2 abatement targets and the effects of climate change, which include more frequent severe weather events. Despite the cap on carbon emissions, natural gas remains a crucial component in the transition of the power mix, providing 121 TWh of electricity, one-third of the total annual power supply. Our analysis underscores the importance of addressing curtailment, particularly in extreme weather scenarios where excess electricity generation reaches 5 TWh. Although this constitutes a small fraction of total demand, it presents opportunities for integration into hard-to-abate sectors. It highlights the potential for load shifts to manage electricity demand more efficiently. The projected increase in annual electricity demand, from 320 TWh in 2019 to 331-334 TWh in 2030, reflects the anticipated electrification of end-uses, balanced by investments in energy efficiency. However, this aggregate increase masks significant hourly variations, especially during summer peaks.

Finally, this study underscores the vital interplay between mitigation and adaptation in the power sector. The analysis reveals critical insights into the costs and benefits associated with achieving mitigation goals in the context of climate change impacts on the Italian power system. We compare scenarios with and without stringent mitigation targets (denominated $Non Mitigated$ and $Mitigated$ cases), considering the set of different meteorological scenarios deployed. In the $Mitigated$ cases, we demonstrate that, while initial investments in renewable technologies are higher, they significantly reduce operational costs associated with fuel purchases. This shift from operational to capital expenditures enhances energy independence and resilience to geopolitical frictions and price volatility. The increase in capital investments and the decrease in operational and fuel purchase expenditures requires a careful evaluation of the financial aspects linked to this shift, which falls out of the scope of this work. We leave for further analysis the assessment of how the change in these financial flows affects the actors of the supply chain, as well as the possible regulatory incentives for the new capital investments; both aspects are crucial for achieving an effective energy transition.
Solar panels emerge as a pivotal technology in the trade-off between adaptation and mitigation of the power system, contributing to environmental sustainability and bolstering system resilience against rising temperatures. Notably, the study highlights that the total annual system costs remain relatively stable between the $Non Mitigated$ and $Mitigated$ scenarios, underscoring the feasibility and advantages of concurrently pursuing decarbonization and climate adaptation strategies. These findings underscore the critical importance of integrating renewable energy investments into long-term planning to develop a robust, sustainable, and climate-resilient power system.

%Finally, another key message of this work concerns the synergies between mitigation and adaptation in the power sector. Integrating adaptation measures into decarbonization scenarios results in a slightly greater cost escalation compared to integrating such measures in a scenario without climate policies. The effects of climate impacts, though, increase the load and reduce the availability of some power sources. When no mitigation policy is enforce an extreme weather year in the future may yield to a non negligible increase in emissions. The need for transition policy importance is thus exacerbated when considering the possible negative impacts of extreme weather and climate change. Adapting in a mitigated power sector necessitates amplified capital expenditure investments compared to heightened operational expenditures prevalent in scenarios lacking mitigation efforts. Undoubtedly, this poses significant challenges concerning both funding and public acceptance. The shift toward this approach signifies a fundamental restructuring of investment strategies, emphasizing upfront capital outlays.

\section{Conclusions}
\label{sec: conclusions}

Our findings indicate that transitioning to a low-carbon power system in Italy by 2030, capable of meeting demand under increasingly extreme weather conditions, requires substantial investments in PV generation and storage capacity. Utility-scale PV is identified as the most cost-efficient option, with rooftop PV playing a crucial role in alleviating grid stress. Furthermore, resilience to extreme weather events necessitates a robust storage strategy, with lithium-ion batteries preferred for short-term energy management due to their round-trip efficiency. Moreover, our results display the projected generation mix for 2030, showing solar electricity production increasing substantially due to stringent CO$_2$ abatement targets and adaptation to impacts of climate change. Despite the emphasis on renewable energy, natural gas still supplies one-third of total annual electricity, while managing curtailment is crucial for redirecting excess generation to hard-to-abate sectors. Finally, this study highlights the crucial interaction between mitigation and adaptation strategies in the power sector, demonstrating that while initial investments in renewable technologies are higher under stringent mitigation scenarios, they lead to significant reductions in operational costs and enhance the robustness of the system to climate change impacts. Importantly, the findings indicate that total annual system costs remain stable between scenarios with and without stringent mitigation targets, affirming the feasibility of simultaneously pursuing decarbonization and climate adaptation. These results underscore the necessity of integrating renewable energy investments, particularly on solar panels, within long-term planning to develop a sustainable and climate-resilient power system, given their low dependence on meteorological conditions.

This analysis is not without caveats. The impacts of mean climate shifts and extreme events to the transmission lines (i.e. due to overheating of the cables) are not included. In addition, we do not perform any sensitivity analyses on economic variables, which could shift technical choices on different options. This can be a crucial consideration for the resilience of transition scenarios; however, it does not constitute the primary focus of this study.

Future research can expand the evidence provided in this work in several directions. A more detailed power dispatch model could investigate options to strengthen the resilience of power systems not only through changes in the dispatch mix, but also by exploiting balancing services, cross-border trade and demand-side management. Furthermore, given that conventional generation technologies play a dominant role in setting wholesale prices as they meet the net-load, i.e. residual demand not satisfied by renewable sources, extreme weather events may result in wholesale price fluctuations. Understanding the characteristics of power markets’ operations during extreme weather may bring to the surface possible limitations of the current power systems, leading not only to volatility in power prices but possibly also to higher costs for managing the grid. Models with a detailed representation of the use of electric appliances and of the behavioural aspects of consumption can be adopted to investigate the demand-side potentials for reducing the peak-load during extreme events.\\

\section{Acknowledgements}
A.D.B has received support from the GRINS (PNRR) project; F.C. has received support from the DIGITA (PRIN) project.
We acknowledge financial support from the Italy's National Recovery and Resilience Plan (PNRR), grant agreement No PE0000018 - GRINS – Growing Resilient, INclusive and Sustainable.\\

\textbf{CRediT authorship contribution statement}

A. Di Bella and F. P. Colelli: Conceptualization, Data curation, Formal analysis, Investigation, Methodology, Software, Visualization, Writing – original draft, Writing – review \& editing.
\\

\textbf{Declaration of generative AI and AI-assisted technologies in the writing process}
\noindent
During the preparation of this work the authors used ChatGPT in order to improve the language of this paper. After using this tool, the authors reviewed and edited the content as needed and take full responsibility for the content of the publication.

\begin{appendices}
\renewcommand{\theequation}{\thesection.\arabic{equation}}

% reset the counter
\setcounter{equation}{0}

\section{Supplementary Results}

\subsection{Effects of rooftop PV generation on centralised power demand}
\label{appendix: rooftop PV}

\begin{figure}[H]
\begin{center}
\includegraphics[scale=0.5]{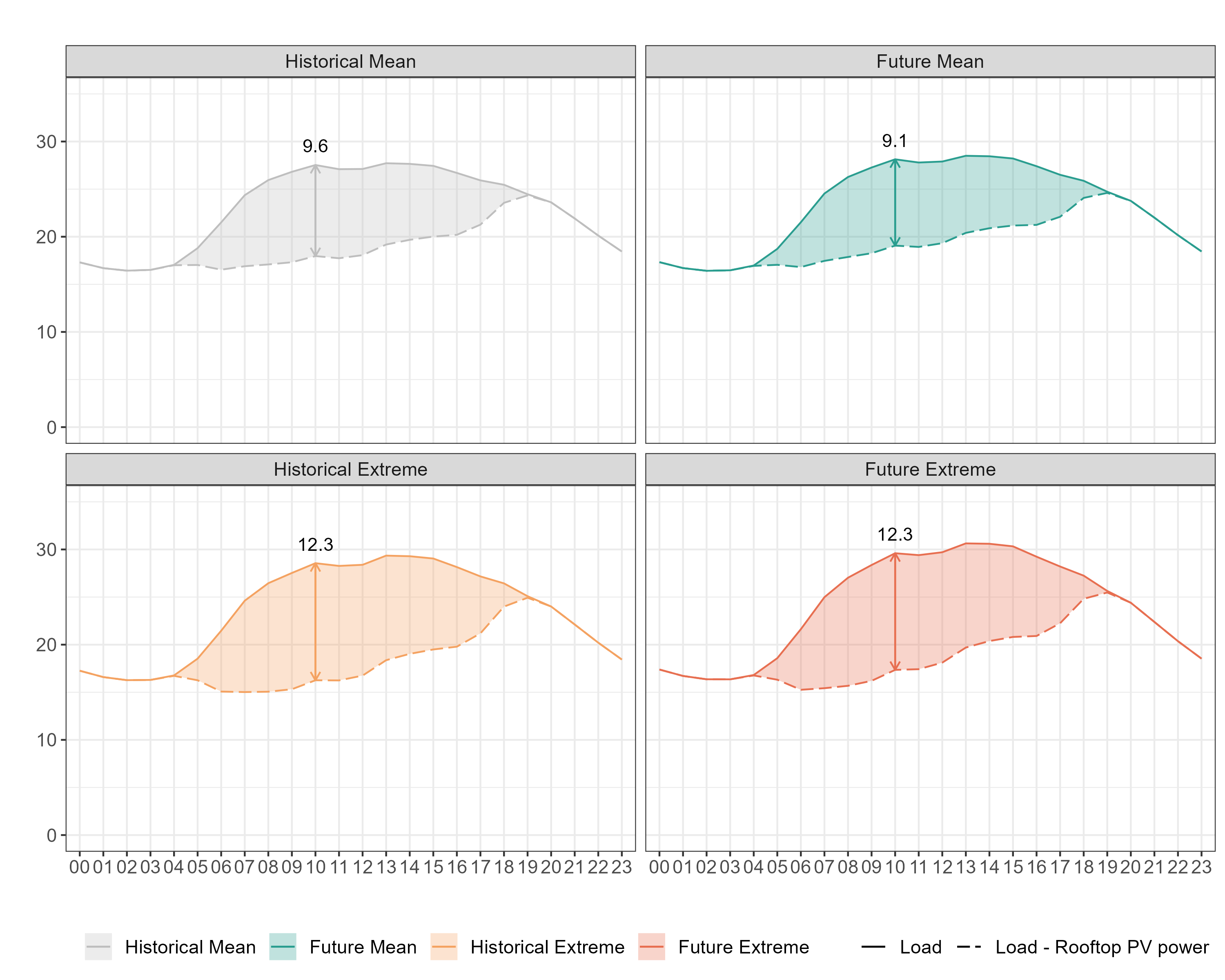}
\caption{Net electricity demand after the one supplied by the distributed generation through rooftop photovoltaic panels, for the four scenarios. The values are for an average of all the Italian regions during the summer season}\label{Fig: net demand after PV}\label{Fig:solar_net_load}
\end{center}
\end{figure}

Rooftop PV generation is a form of distributed power generation, thus it has the advantage of reliving tension from the high voltage grid by producing locally and decreasing the centralised electricity load. In this sense, even if the overall electricity demand increases due to climate change or severe weather conditions, the remaining power request from centralised technologies is reduced, as illustrated in Figure \ref{Fig: net demand after PV}). In the two Extreme scenarios the power load is almost halved by the large presence and generation provided by rooftop PV panels. This technology represents a pivotal resource within mitigation strategies, not only to address increased household electricity consumption, particularly for air conditioning, stemming from climate change-induced warmer climates, but also to alleviate strain over centralised power generation and transmission.

\subsection{Mitigation and adaptation trade-off}
\label{appendix: mitig adapt}

\begin{figure}[H]
\begin{center}
\includegraphics[scale=0.75]{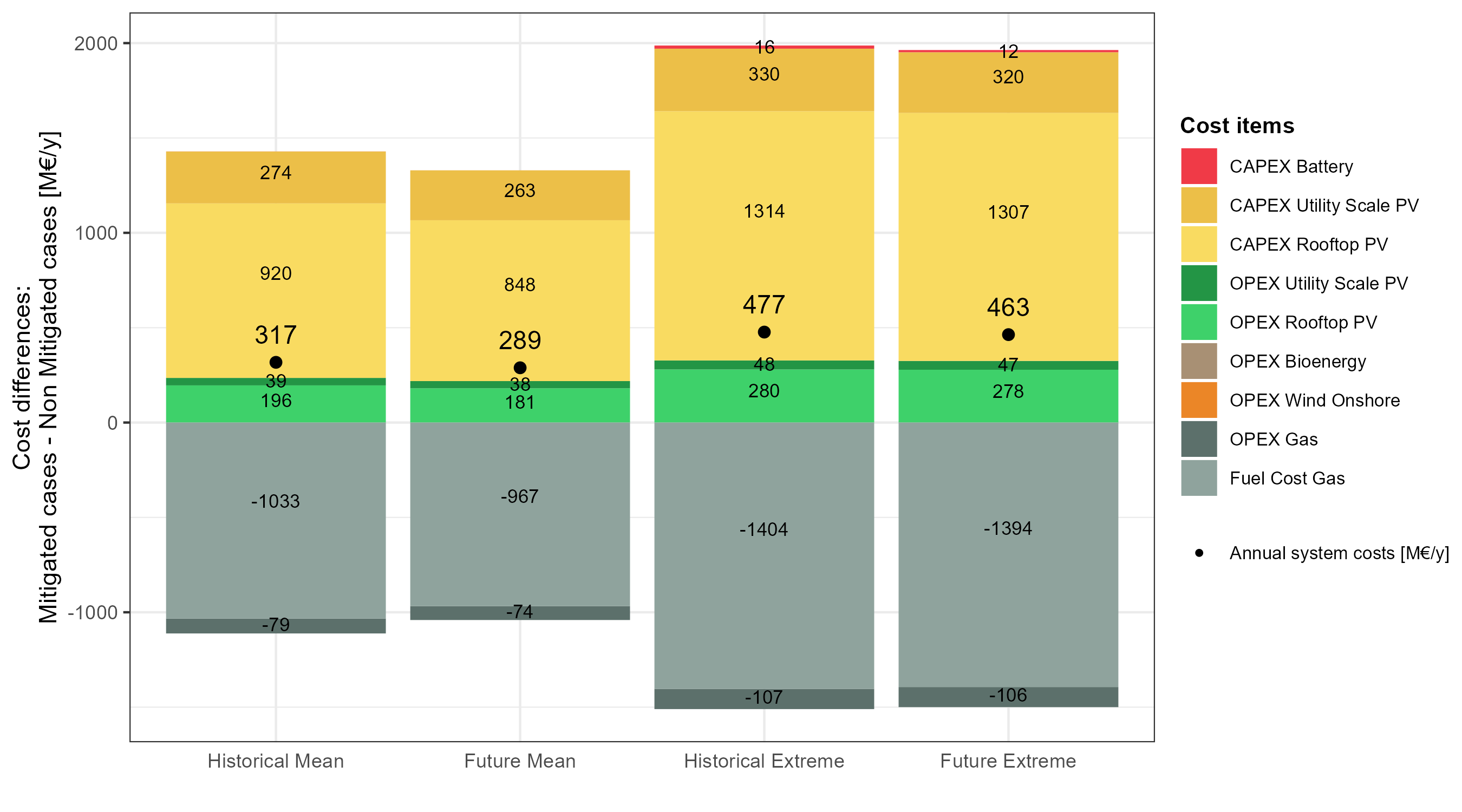}
\caption{Cost items differences in the total balance of the annual system costs for the Italian power system in 2030, $Mitigated$ cases with respect to $Non Mitigated$ cases. Bars represent cost items differences between $Mitigated$ cases and $Non Mitigated$ cases. The black dots indicates the total annual system cost difference between $Mitigated$ cases and $Non Mitigated$ cases.}\label{Fig: cost diff mit no mit}
\end{center}
\end{figure}

Figure \ref{Fig: cost diff mit no mit} represents the additional (if positive) or reduced (if negative) costs that the system incurs under increased weather stress in the mitigated scenarios compared to the scenario without any imposed climate goals. The black dots represent the difference in total annual system costs between the $Mitigated$ and $Non Mitigated$ scenarios. Their presence on the positive side of the graph highlights that achieving both adaptation resilience and decarbonization goals leads to a greater increase in annual power system costs compared to a system addressing only adaptation constraints. The delta in annual system expenditures decreases when looking at the corresponding Future cases (\textit{Future Mean} with respect to \textit{Historical Mean}, \textit{Future Extreme} with respect to \textit{Historical Extreme}). This indicates that adapting to a changing weather in terms of AC demand, thermal and hydro power generation unavailability, reduces the mitigation costs. Solar panels play a key role in this trade-off, as they not only represent an environmentally friendly technology but also enhance system resilience to rising temperatures.

Another critical observation from Figure \ref{Fig: cost diff mit no mit} is that, as anticipated, all the mitigated scenarios exhibit higher expenditures for PV panels and batteries, while reducing costs associated with natural gas acquisition. This has relevant implications in terms of energy security: in fact, the Italian power system in all mitigated cases would be more independent from energy imports.

\begin{figure}[H]
\begin{center}
\includegraphics[scale=0.75]{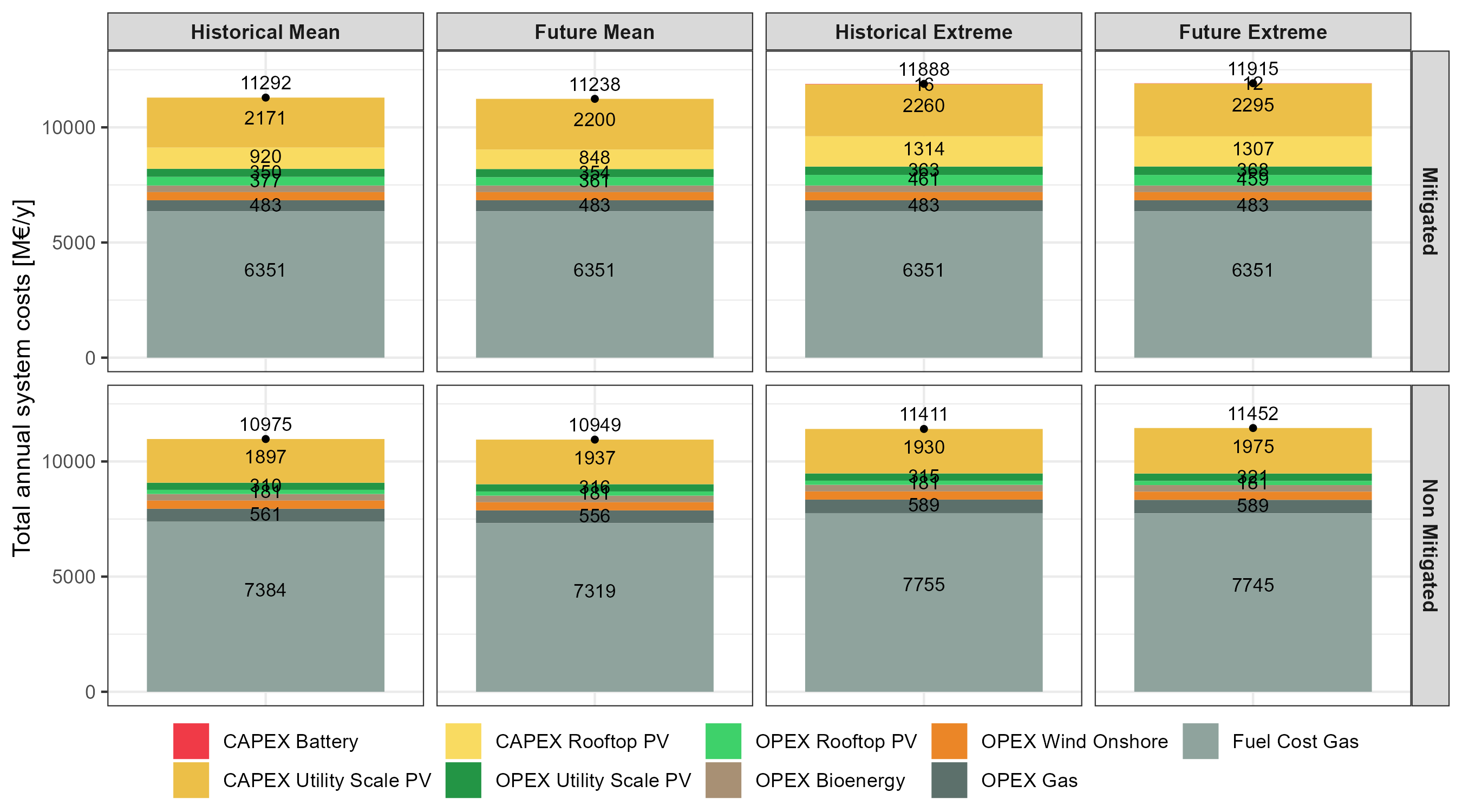}
\caption{Bars represent cost items differences between $Mitigated$ cases and $Non Mitigated$ cases. The black dots indicates the total annual system cost difference between $Mitigated$ cases and $Non Mitigated$ cases.}\label{Fig: cost diff mit no mit absolute}
\end{center}
\end{figure}

Another graph the we deem important to show is Figure \ref{Fig: cost diff mit no mit absolute}. It offers a comprehensive overview of the cost items embedded in the different total annual system costs, allowing to really get the idea of how the expenditures for the Italian power sector in 2030 would be allocated. A notable point it the absence of CAPEX expenses for rooftop PV in the $Non Mitigated$ cases. This distributed technology has a larger cost with respect to Utility Scale solar generation since it lacks economies of scale. It becomes critical though, when achieving mitigation goals, to enhance the production of renewable electricity and ensure the operation of the power system when higher temperatures have an impact on hourly AC demand.

\section{Supplementary Methods for Climate Impacts}
\label{appendixB}
\subsection{Air Conditioning load estimation}
\label{subsubsec: empirical analysis}

 Coefficients are estimated by \cite{colelli2023} though a pooled cross section-time series regression with a logit link function, $\Lambda$ \cite{ferrari2004beta}. The variables used to identify the level of AC adoption are the 10-year moving average CDD24s ($\mathcal{C}$), the logarithm of the 10-year moving average annual per capita income ($y$) and the logarithm of the 10-year moving average annual urbanization rate ($u$):
\begin{align}
\Lambda (s_{i,t}) &= \log \left( \frac{s_{i,t}}{1-s_{i,t}} \right) = \boldsymbol{Z}\boldsymbol{\alpha} \nonumber\\
&= \alpha_i^0 + \alpha^Y y_{i,t}  + \alpha^C \mathcal{C}_{i,t} + \alpha^{YC} (y_{i,t} \cdot \mathcal{C}_{i,t}) + \alpha^U u_{i,t} \label{eq:eq_ac_adoption} 
\end{align}
with location fixed effects $\alpha^0$, and estimated parameters $\alpha^Y$ and $\alpha^C$ that capture the direct effects of income and heat exposure, and $\alpha^{YC}$ that captures their interaction. The functional form yields nonlinear effects of the linear predictors, governed by the logistic transformation.

The non-linear temperature-load function is estimated with a fixed effect models of per capita daily electric load, $q$, at the European member state level in each day from 2015 to 2019. Population-weighted temperatures are binned into $k$ intervals of 3$^\circ$C width, $B_k=[\underline{T}_k, \overline{T}_k)$ to construct a $k$-vector of indicators that track whether each day's maximum temperature falls within a given interval:
$$\mathcal{T}_k = 1 \cdot \lbrace T \in B_k \rbrace + 0 \cdot\lbrace\text{Otherwise}\rbrace$$. Suppressing location and time subscripts, the empirical specification estimated by \cite{colelli2023} is:
\begin{align}
\mathbb{E} [\ln q] &=  \textstyle \sum_k \beta_{k,v}^T \mathcal{T}_k +   \sum_k \beta_{k,v}^{TAC} \left( \mathcal{T}_k \cdot s \right) + \beta_v^Y y + \text{controls}  \label{eq:ac_interaction} \end{align}
where controls include state or country fixed effects that absorb variation associated with unobserved temporally-invariant confounders, and day-of-week, season and year fixed effects that control for idiosyncratic time-varying influences that are unrelated to temperature. The elements of $\boldsymbol{\beta}^T$ account for consumers' adjustments of stocks of energy-using durables, explicitly captured by the vector of interaction coefficients, $\boldsymbol{\beta}^{TAC}$. The fitted coefficient vectors $\widehat{\boldsymbol{\beta}}^T$ and $\widehat{\boldsymbol{\beta}}^{TAC}$ provide flexible piece-wise linear spline representations of macro-regions' distinct nonlinear temperature response functions (see Supplementary Information). 

\subsection{Impacts on thermal generation outages}
\label{subsec: Thermal generation outages}

We study the potential unavailability of thermal power generation units due to extreme temperatures by developing a regression model based on outage information collected from 2018 to 2022 in Italy. The dataset \cite{entsoe_outages} includes information of over 4000 outages, of which 2352 of gas- and 1887 of coal-fired generation units. The method adopted partially follows previous analysis (\cite{coffel2021thermal}), but expands from the literature by providing country-specific and fuel-specific responses. We consider as key dependent variable the occurrence of an outage in each power-plant, represented by a dichotomous variable alternatively 0 or 1 ($c$), and identify the influence of daily maximum temperatures ($t$) and water runoff anomalies ($r$), controlling for month (m) and location (p) fixed effects. Since temperature and runoff anomalies can have non-linear impacts on the operations of thermal generators, non-linear effects are captured by adding a quadratic or alternatively a cubic term to each. Furthermore, in the most detailed specification, we identify fuel-specific impacts by including a set of interaction terms between temperature and runoff anomalies and a character variable representing the power plant fuel $k$. We estimate the following equation though a logistic regression, so that the log odds of the outcome are modelled as a combination of the predictor variables:

\begin{align}
\Lambda (c_{i,t}) &= \log \left( \frac{c_{i,t}}{1-c_{i,t}} \right) = \emph{f} (t_{i,t}) \cdot k + \emph{z} (r_{i,t}) \cdot k + \psi k + \mu p + \nu m + \varepsilon\label{eq:outages} \\
\end{align}

\begin{figure}[H]
\begin{center}
\includegraphics[scale=0.5]{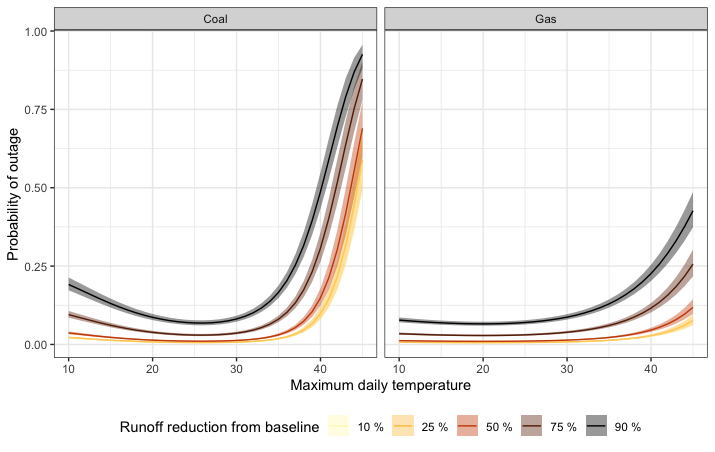}
\caption{Outage probability based on temperature and water runoff anomalies levels, based on the selected logit model. Shades show the 95th confidence interval.}\label{Fig:outage_models}
\end{center}
\end{figure}

We compare the different polynomial specifications based on standard performance metrics, and select the model with a fuel-specific cubic response to temperatures (Figure \ref{Fig:outage_models}). 

We find that the likelyhood of occurrence of an outage for coal-fired generation increases considerably when daily maximum temperatures surpass 35°C, reaching 10-50\% at 40°C and 50-90\% at 45°C, depending on the water runoff anomaly. Gas-fired generation is less sensitive to high temperature and low water runoff levels, but the likelihood of an outage is non-negligible and between 5-25\% at 40°C. Regression results are shown in the Table \ref{tab:outagreg}. We use the estimated available capacity function in conjunction with future daily maximum temperature anomalies to simulate daily thermal power generation availability around 2030 and apply the results to the oemof Italian power system model optimisations.

\begin{table}[!htbp] \centering 
  \caption{Outage regression results} 
  \label{tab:outagreg} 
\begin{tabular}{@{\extracolsep{5pt}}lccc} 
\\[-1.8ex]\hline 
\hline \\[-1.8ex] 
 & \multicolumn{3}{c}{\textit{Dependent variable:}} \\ 
\cline{2-4} 
\\[-1.8ex] & \multicolumn{3}{c}{Likelihood of outage} \\ 
\\[-1.8ex] & (1) & (2) & (3)\\ 
\hline \\[-1.8ex] 
 technology-Gas & $-$1.760$^{***}$ & $-$2.471$^{***}$ & $-$0.973$^{***}$ \\ 
  & (0.050) & (0.073) & (0.123) \\ 
  & & & \\ 
 tmax & $-$0.037$^{***}$ & $-$0.244$^{***}$ & 0.103$^{***}$ \\ 
  & (0.002) & (0.006) & (0.019) \\ 
  & & & \\ 
 tmax$\hat{\mkern6mu}$2 &  & 0.005$^{***}$ & $-$0.015$^{***}$ \\ 
  &  & (0.0001) & (0.001) \\ 
  & & & \\ 
 tmax$\hat{\mkern6mu}$3 &  &  & 0.0003$^{***}$ \\ 
  &  &  & (0.00002) \\ 
  & & & \\ 
 runoff\_anomaly & $-$0.758$^{***}$ & $-$0.482$^{***}$ & $-$0.370$^{***}$ \\ 
  & (0.065) & (0.065) & (0.066) \\ 
  & & & \\ 
 runoff\_anomaly$\hat{\mkern6mu}$2 & 4.059$^{***}$ & 3.638$^{***}$ & 3.536$^{***}$ \\ 
  & (0.074) & (0.075) & (0.075) \\ 
  & & & \\ 
 technology-Gas:tmax & 0.063$^{***}$ & 0.155$^{***}$ & $-$0.129$^{***}$ \\ 
  & (0.001) & (0.006) & (0.020) \\ 
  & & & \\ 
 technology-Gas:tmax$\hat{\mkern6mu}$2 &  & $-$0.002$^{***}$ & 0.013$^{***}$ \\ 
  &  & (0.0002) & (0.001) \\ 
  & & & \\ 
 technology-Gas:tmax$\hat{\mkern6mu}$3 &  &  & $-$0.0003$^{***}$ \\ 
  &  &  & (0.00002) \\ 
  & & & \\ 
 Constant & $-$4.036$^{***}$ & $-$2.552$^{***}$ & $-$4.271$^{***}$ \\ 
  & (0.075) & (0.085) & (0.125) \\ 
  & & & \\ 
  month & Y & Y & Y\\ 
  year & Y & Y & Y\\ 
  province & Y & Y & Y\\ 

\hline \\[-1.8ex] 
Observations & 1,353,432 & 1,353,432 & 1,353,432 \\ 
Log Likelihood & $-$170,880.300 & $-$170,056.200 & $-$169,796.300 \\ 
Akaike Inf. Crit. & 341,854.700 & 340,210.400 & 339,694.600 \\ 
\hline 
\hline \\[-1.8ex] 
\textit{Note:}  & \multicolumn{3}{r}{$^{*}$p$<$0.1; $^{**}$p$<$0.05; $^{***}$p$<$0.01} \\ 
\end{tabular} 
\end{table} 

\subsection{Impact on variable renewable sources}
\label{subsubsec: vre impacts}

We describe here the methodology for assessing the impacts of climate change on variable renewable generation production. We only evaluate these impacts in the two scenarios \textit{Historical Mean} and  \textit{Future Mean}, since the optimization of power system capacity expansion is not be based on extreme low- or high- wind and solar generation occurrences, which are typically dealt with by transmission system operators though ancillary services and balancing energy markets. In other words, we only consider how climate change may alter the solar and wind production patterns by taking into account the hour-, calendar day- and region- specific mean production value across the historical and future year period. For both wind and solar power potential generation, we exploit the projections developed by \cite{secures}. For wind generation, we retrieve the time-series of normalised with potential at the NUTS3 level directly from the database, then we aggregate at the bidding-zone level. As for solar, we consider the time-series of hourly population-weighted direct normal irradiation (DNI), the amount of solar radiation per unit area [W/m$^2$] by a surface perpendicular to the sun rays. Temperature is also included into the analysis, since the power output that a solar panel \textit{p} is able to produce at each hour \textit{h} in its location \textit{n} is dependent on ambient temperature. The following formula, taken from \cite{noauthor_ieee_nodate} shows the correlation, where T$_{n,h}$ is expressed in degree Celsius and DNI$_{n,h}$ is the DNI value, in location \textit{n} and hour \textit{h}:

\begin{equation}
  P_{n,h}  = \eta_{p} S_{p}  DNI_{n,h} (1-0.005(T_{n,h}-25))
\end{equation} 

The parameters $\eta_{p}$ and $S_{p}$ represent the conversion coefficient [\%] and the surface area [m$^2$] specific to the photovoltaic unit \textit{p}. These values are not relevant for the calculation since $P_{n,h}$ is then normalised to a time-series, accounting for the maximum values in each scenario and region, to obtain the solar potential. $\eta_{p}$ itself is dependent on temperature, as explained by Evans \cite{evans_simplified_1981}, following the equation:

\begin{equation}
  \eta_{p,T}  = \eta_{p,Tref} * (1 - \beta_{ref} * (T - Tref))
\end{equation} 

$\eta_{p,Tref}$ refers to the panel efficiency at Tref, the reference temperature (which is commonly 25°C) and solar radiation of 1000 W/m$^2$. $\beta_{ref}$ is the temperature coefficient and it is generally assumed to be 0.0004 K$^{-1}$ \cite{dubey_temperature_2013,notton_modelling_2005}. Thanks to these equations  we can coherently consider the impact of the change of DNI and temperature in the future climate scenarios on PV panels power output.

\begin{figure}[H]
\begin{center}
\includegraphics[scale=0.5]{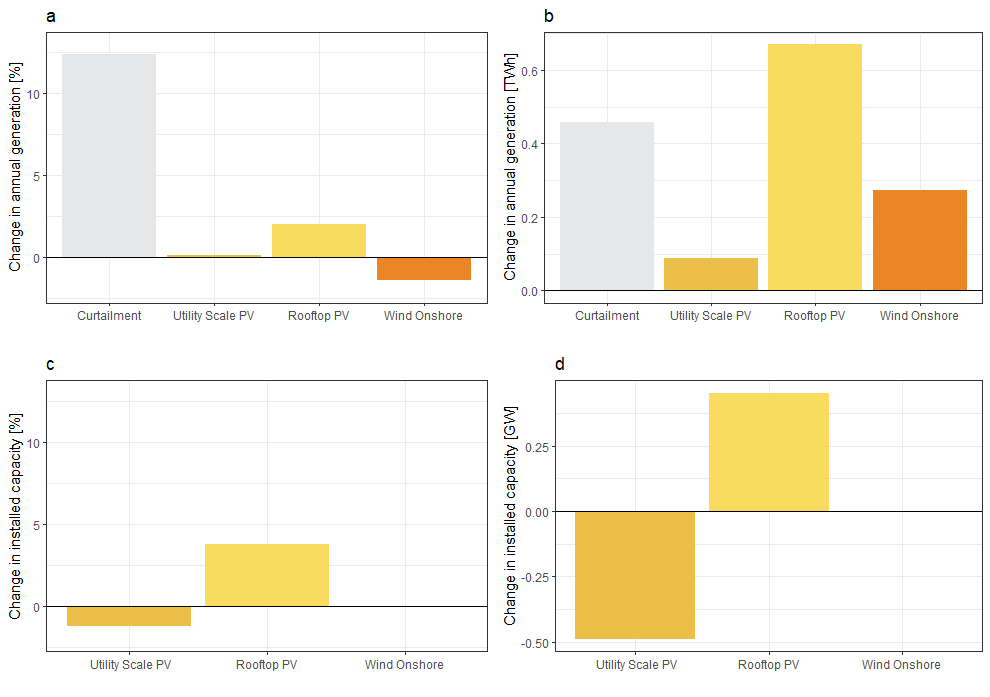}
\caption{Change in the generation (panels a-b) and capacity (panels c-d) in the $Future Mean$ climate scenario when including the mean climate change impacts on the hourly wind and solar generation.}\label{Fig:impact_res}
\end{center}
\end{figure}

In Figure \ref{Fig:impact_res} we evaluate the variation in electricity generation and additional installed capacities for the Italian case study in 2030 in the \textit{Future Mean} weather scenario with respect to the \textit{Historical Mean} one. Again, we do not focus on the \textit{Historical Extreme} and \textit{Future Extreme} scenario because the high variability of renewable sources to weather is informative for dispatch optimisation problems, while it is not a key driver for power system investment planning. The input shocks included in the model, namely the mean level of the normalised wind power potential and the BNI, used to project future impacts for wind and solar, respectively, are presented in Figure \ref{Fig:generation_impacts}, Panels c and d. As you can see in Figure \ref{Fig:impact_res}, the changes

\subsection{Additional figures}
\label{subsubsec: additional figures}

\begin{figure}[H]
\begin{center}
\includegraphics[scale=0.5]{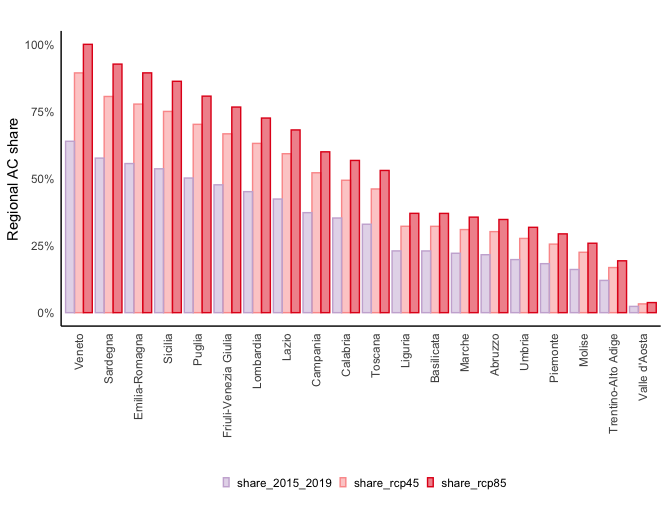}
\caption{\footnotesize AC prevalence in households in Italian regions, observed average of 2015-2019, and future in 2030 under RCP 4.5 and RCP8.5.}\label{Fig:regional_ac_share}
\end{center}
\end{figure}

\begin{figure}[H]
\begin{center}
\includegraphics[scale=0.5]{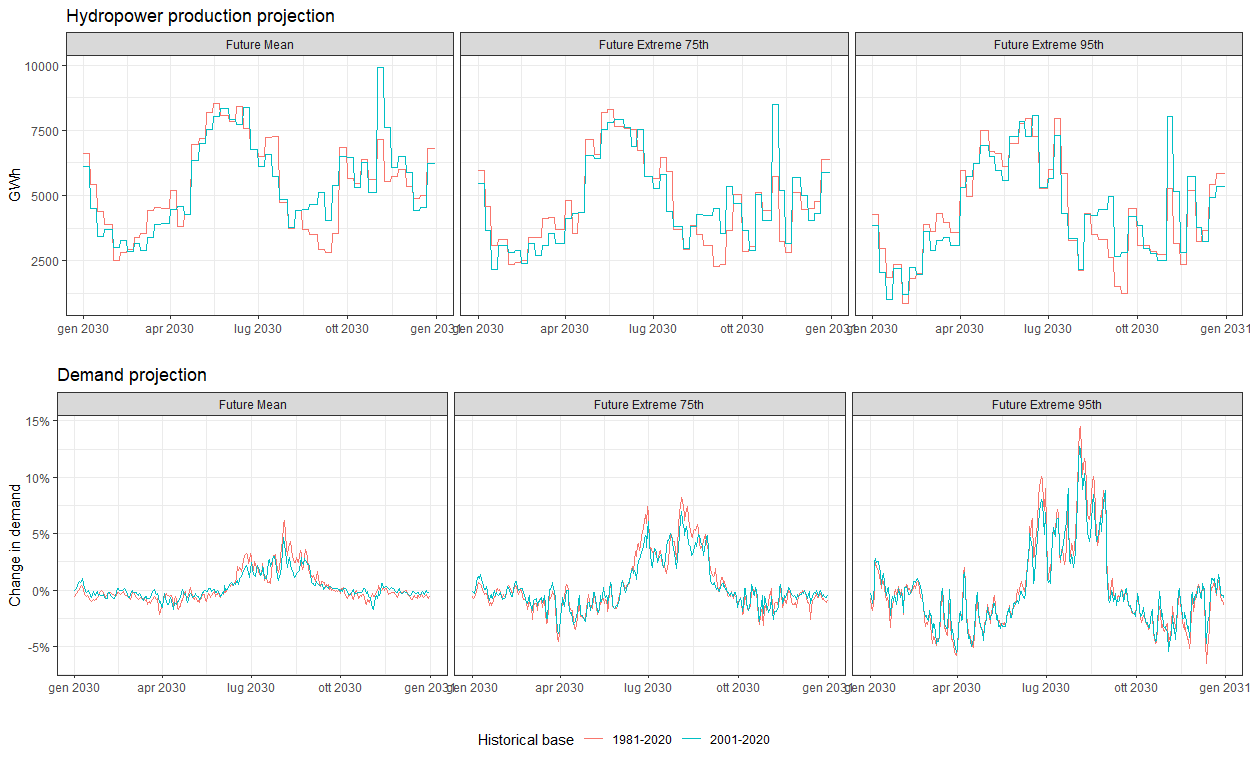}
\caption{\footnotesize Comparison of input projections of hydro production and electricity demand change, by choice of historical base year period}\label{Fig:base_year_comparison}
\end{center}
\end{figure}

Figure \ref{Fig:base_year_comparison} shows that no substantial difference can be found in the value of the input projections of hydro production and electricity demand change by changing historical base year period of CMIP6 model output from 1981-2020 to 2001-2020.

\begin{figure}[H]
\begin{center}
\includegraphics[scale=0.5]{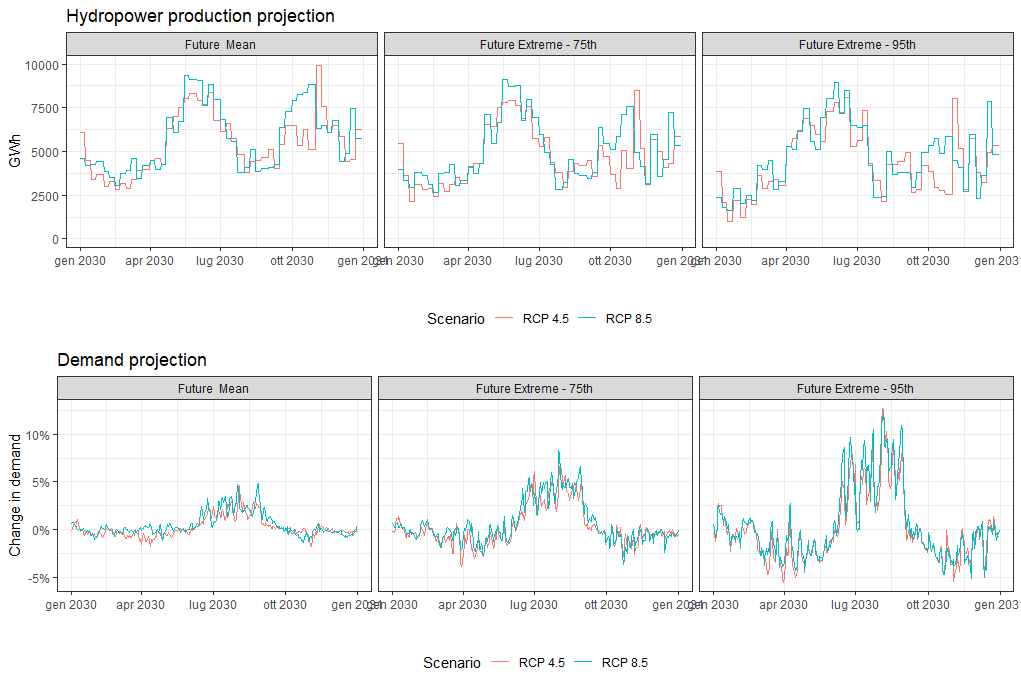}
\caption{\footnotesize Comparison of input projections of hydro production and electricity demand change, by RCP}\label{Fig:rcp_comparison}
\end{center}
\end{figure}

Figure \ref{Fig:rcp_comparison} shows that no substantial difference can be found in the value of the input projections of hydro production and electricity demand change from RCP 4.5 and RCP 8.5 in 2030.

\section{Supplementary Methods for Power System Modelling}
\label{appendixC}
\setcounter{equation}{0}

We employ a power system model of the Italian electricity system \cite{Di_Bella_Oemof_Italy_2022}, developed with Oemof (Open Energy MOdelling Framework), an open-source energy modelling tool in Python \cite{oemof}. This framework enables the creation of an energy network and then uses a solver to determine the energy balances (in this case, the one utilized was Gurobi, but Oemof is able to work with other open-source solvers). The same power system model for Italy has been used in previous works \cite{di_bella_power_2024, di_bella_demand-side_2022}. To be able to run simulations and investment optimizations, Oemof expansion capacity script requires the following inputs, provided through an Excel file. The spatial resolution of the model is based on the Italian electricity market zones, which, since year 2021, consist of 7 different areas \cite{Terna7zones, Terna7zones_allegato}, organized as shown in Table \ref{tab:marketzonesbreakdown}. Building from the subsection in the paper discussing briefly the main elements of the model, Table \ref{tab:marketzonesbreakdown} shows the break-down in regions of each model node, while Table \ref{tab: transmission lines} outlines the powerlines capacity.

\begin{minipage}[b]{0.40\linewidth}
\begin{table}[H]
    \centering
    \footnotesize
    \setlength\extrarowheight{3pt}
    \caption{Italian electricity market zones composition}
    \begin{tabular}{m{8.8em}  m{12em}}
    \hline
        \textbf{Market zone}&\textbf{Regions}\\
        \hline
        North & Emilia-Romagna, Friuli Venezia-Giulia, Liguria, Lombardia, Piemonte, Trentino-Alto Adige, Valle D'Aosta, Veneto\\
        Centre-North& Marche, Toscana, Umbria \\
        Centre-South& Abruzzo, Campania, Lazio \\
        South& Basilicata, Calabria, Molise, Puglia \\
        Sardinia& Sardinia \\
        Sicily& Sicily  \\
        Calabria& Calabria \\
    \hline
    \end{tabular}
    \label{tab:marketzonesbreakdown}
\end{table}
\end{minipage}\hfill%
% CONTROLLARE I DATI IN BASE ALL'ANNO
\begin{minipage}[b]{0.50\linewidth}
\begin{table}[H]
    \centering
    \footnotesize
    \caption{Transmission lines capacity}
    \begin{tabular}{c c}
    \hline
    \textbf{Line}&\textbf{Capacity [MW]}\\
    \hline
        North $\xrightarrow{}$ Centre-North& 4300 \\
        North $\xleftarrow{}$ Centre-North& 3100 \\
        \hline
        Centre-North $\xrightarrow{}$ Centre-South& 2900 \\
        Centre-North $\xleftarrow{}$ Centre-South& 2800 \\
        \hline
        Centre-South $\xrightarrow{}$ South& 2000 \\
        Centre-South $\xleftarrow{}$ South& 5550 \\
        \hline
        Centre-South $\xrightarrow{}$ Sardinia& 720 \\
        Centre-South $\xleftarrow{}$ Sardinia& 900 \\
        \hline
        Centre-North $\xrightarrow{}$ Sardinia & 1095 \\
        Centre-North $\xleftarrow{}$ Sardinia & 1015 \\
        \hline
        Sicily $\xrightarrow{}$ Sardinia& 800  \\
        Sicily $\xleftarrow{}$ Sardinia& 800  \\
        \hline
        Centre-South $\xrightarrow{}$ Sicily& 700  \\
        Centre-South $\xleftarrow{}$ Sicily& 700  \\
        \hline
        South $\xrightarrow{}$ Calabria& 1100\\
        South $\xleftarrow{}$ Calabria& 2350 \\
        \hline  
        Calabria $\xrightarrow{}$ Sicily& 1750 \\
        Calabria $\xleftarrow{}$ Sicily& 1200 \\
    \hline
    \end{tabular}
    \label{tab: transmission lines}
\end{table}
\end{minipage}\hfill%

\subsection{Key features}

\begin{figure}[H]
\begin{minipage}[b]{0.5\linewidth}
\centering
\includegraphics[scale=0.5]{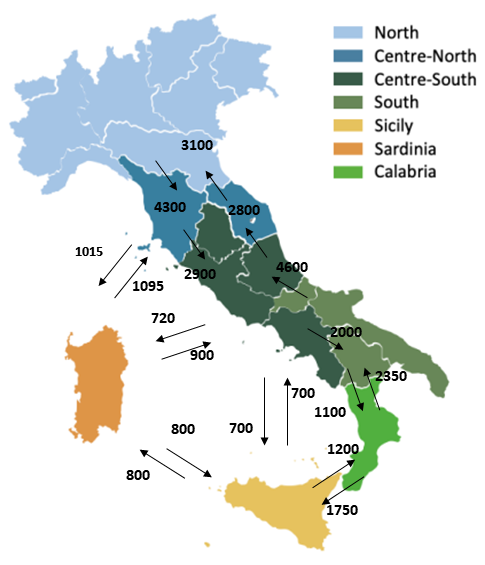}
\end{minipage}
\begin{minipage}[b]{0.5\linewidth}
\caption{Representation of the spatial resolution of the Italian power model. The seven nodes of the model aggregate different Italian regions, according to the colours in the picture. The zones are linked through high-voltage power lines, whose transmission capacity is indicated with the values in the graph [GW]. The capacity values assume that the already announced grid expansion projects up until 2026 are built \cite{powerlinesterna21}.}
\label{Fig:italian_peninsula}
\end{minipage}
\end{figure}

\textit{Transmission lines}. Market zones are linked in the model through high-voltage transmission lines. Capacity data are taken from Terna Development Plan in 2019 \cite{powerlinesterna} and updated according to the 2021 Plan \cite{powerlinesterna21} and the National Energy and Climate Plan (PNIEC \textit{Piano Nazionale Integrato per l'Energia e il Clima} \cite{PNIEC}), assuming the  transmission capacities targets for 2026 completely achieved in 2030. The spatial subdivision and the values for exchangeable GWs between zones are visualised in Figure \ref{Fig:italian_peninsula}.

\textit{Demand}. For each region, a time series of the hourly power load is considered, obtained from Terna \textit{Download Center} web page, taking data for the year 2021  \cite{modelvalidationterna}. Terna projects an annual national electricity demand of 331 TWh by 2030, taking into account the ongoing electrification of the transportation, heating, and industrial sectors \cite{TernaAdeguatezza}. The distribution of this demand into market zones in 2030 is assumed to mirror that of 2021. Consequently, for 2030, we maintain the same hourly demand profiles, adjusting them upward to align with the anticipated national load of 331 TWh for the year.\\

\textit{Commodity sources}. For power production, the resources considered in this case study are natural gas, water and imported electricity. Nowadays Italy still has some coal-fired power plants  running \cite{AssoCarboni}, but according to the Italian climate and energy strategy \cite{PNIEC}, the phase-out from this polluting source will be in 2025. Therefore, this work assumes that in 2030 there will be no coal power plant working. Import refers to the net electricity imported from other countries and it has been designed as a source of generation. Italy exchanges electricity with France, Switzerland, Austria, Slovenia, Greece, Malta and Montenegro, but the main importer is France \cite{Terna7zones_allegato}. For this reason we assumed to model import as a source of generation only for the \textit{North} market zone and with a negligible emission factor \cite{EmissionFactors}, thanks to the large fraction of French electricity produced with nuclear power. For commodity sources, variable costs are specified; for import and hydro costs are set to zero. Natural gas prices for 2030 are derived from World Energy Outlook 2022 \cite{WEO2022}, assuming the Stated Policy case, thus 29 €/MWh of thermal energy. Specific emissions factors are adopted from technical specifications in \cite{EmissionFactors}.\\

\textit{Power plants transformers}. For natural gas-fired plants, total installed capacities are imposed at 2021 values \cite{TernaImpianti2021}, with no possibility to expand. Import power derives from the assumption of constant import throughout the whole year, so a total amount of 42.8 TWh \cite{TernaProduzione2021} reduced by 4.75 TWh due to transmission and distribution losses, results in 4.34 GW in each hour. For hydro power plants, efficiencies for the conversion to electricity are offered by \cite{PyPSA-Eur}, import has no efficiency and for gas-fired plants efficiency is evaluated as an average of the whole Italian gas power plants park (55.1\%) \cite{EfficiencyGas}. For gas plants cost of operation is also specified, set at 4 €/MWh \cite{PyPSACosts}.
\\

\textit{Renewables}. The Renewable Energy Sources RES embedded in this study are rooftop and utility scale photovoltaic, wind onshore and offshore, run-off-river, reservoir hydro, biomass and geothermal generation (actually present only in Tuscany, thus with 771.8 MW in Centre-North). Existing offshore capacity is set to zero and the model can expand generation capacity of the two types of solar and the two categories of wind power. In each node the existing installed power in April 2022 is the starting point, aggregated in market zones. Data are taken from Terna \cite{TernaConsistenzaRinnovabili}, assuming that Utility Scale photovoltaic (PV) has a size larger than 1 MW, and from the European Commission's \textit{Joint Research Centre} (JRC) \cite{JRCHydro} for run-off-river and reservoir hydro. PV panels in large fields or on rooftops are distinguished in the model by different existing capacities,  investments and operational costs and maximum potential for available land. The existing capacity for each power generation technology in each region is represented in Table \ref{tab: renewable capacity 2021}.

\begin{table}[H]
    \centering
    \footnotesize
    \caption{Renewable energy sources and fossil fuels power plants capacity per Italian market zone in 2021 [MW]}
    \begin{tabular}{c c c c c c c c}
    \hline
        \textbf{Capacities in MW}&\textbf{Rooftop PV}&\textbf{Ut. Sc. PV}&\textbf{Wind onshore}&\textbf{Biomass}&\textbf{Run-off-river}&\textbf{Reservoir hydro}&\textbf{Gas-fired power plant}\\
    \hline
        North        & 9397  & 1465  & 172 & 2304  & 4399 & 4352 & 27712 \\
        Centre-North & 1864  & 287 & 162 & 452.0 & 368 & 80 & 2587 \\
        Centre-South & 2687 & 1299  & 2189  & 529 & 1117 & 201 & 7485 \\
        South        & 2718  & 900 & 4760  & 490 & 23 & 122 & 4882 \\
        Sardinia     & 504 & 603 & 1112  & 125 & 20 & 88 & 1088 \\
        Sicily       & 1088 & 601 & 2039  & 104 & 8 & 65 & 5394 \\
        Calabria  & 456 & 149 & 1174 & 219 & 121 & 589 & 3630 \\
    \hline
    \end{tabular}
    \label{tab: renewable capacity 2021}
\end{table}

\textit{Storages}.  In 2021, in the Italian power system only Pumped Hydro Storage (PHS) power plants can store electricity, since there are no batteries or hydrogen storage connected to the grid. Values for PHS pumping and generation power capacities and the nominal retainable energy are provided by JRC database \cite{JRCHydro}, with a national storage value of 560 GWh. Other parameters supplied are a capacity loss of 8.33 $\cdot 10^{-6}$ \cite{PRINA2020114728} (referred to the stored energy), inflow and outflow efficiencies, choosing respectively of 85\% and 90\% to get a Round Trip Efficiency (RTE) of 76.5\%, in line with numbers from the \textit{U.S. Department of Energy} \cite{PHSEfficiency}. The optimisation can expand the storage capacity for lithium-ion batteries and hydrogen storage technology, composed of electrolysers, fuel cells and hydrogen tanks. Further details can be found in Appendix \ref{appendixC}.

\subsection{Dispatch optimization}
\label{subsubsubsec: Dispatch optimization}

In this work we deploy both dispatch and expansion capacity optimizations, based on the chosen scenario. The main goal of a dispatch optimization is to find the generation mix which meets the demand at the lowest operational cost. The solver will start using the cheapest sources per MWh until the energy needed is provided. The objective function developed for this work can be expressed by Eq. \ref{eq:min operation cost}, describing the minimum of the cost to operate the system C$^\textnormal{ operation}$.

\begin{equation} \label{eq:min operation cost}
    \textnormal{Min }\textnormal{C}^\textnormal{ operation} = \textnormal{Min}( \sum_{t=1}^{T} \sum_{n=1}^{N} \sum_{s=1}^{S} \textnormal{E}_{t,n,s} \cdot \textnormal{vc}_{t,n,s})
\end{equation}

where:

\textit{t} = analysed time step, from 1 to T

\textit{n} = node considered, from 1 to N

\textit{s} = generation source, from 1 to S

E$_{t,n,s}$ = energy generated by the source \textit{s}, located in the node \textit{n} at the time step \textit{t}.

vc$_{t,n,s}$ = variable costs of generation for the energy source E$_{t,n,s}$\\

This main objective function must respect a series of constraints. They will be explained and analysed further, following Prina et al. work \cite{PRINA2020114728}. The major constraint is to meet the energy demand, considering also the charge and discharge energy of the storages and the transmission losses. This has to be valid for every time step and it is possible to have an excess of generation.

\begin{equation}
\label{eq:energy balance}
    \sum_{s=1}^{S} (\textnormal{E}_{t,n,s} + \textnormal{E}_{t,n,st}^\textnormal{ charge} - \textnormal{E}_{t,n,st}^\textnormal{ discharge} - \textnormal{E}_{t,p}^\textnormal{ transmission loss}) = \textnormal{D}_{t,n} + \textnormal{E}_{t,n}^\textnormal{ excess}
\end{equation}

E$_{t,n,st}^\textnormal{ charge}$ = energy employed for charging storage technology \textit{st} at time \textit{t} in node \textit{n}

E$_{t,n,st}^\textnormal{ discharge}$ = \textnormal{energy employed for discharging storage technology \textit{st} at time \textit{t} in node \textit{n}}

E$_{t,p}^\textnormal{ transmission loss}$ = energy lost due to transport losses at time \textit{t} in powerline \textit{p} for node \textit{n}

D$_{t,n}$ = load demand at time \textit{t} for node \textit{n}

E$_{t,n}^\textnormal{ excess}$ = surplus of generated energy at time \textit{t} for node \textit{n}\\

Another restriction is the respect of the maximum power given as input for each generator unit. Non-dispatchable renewable units \textit{s} in node \textit{n} can supply up to their nominal capacity P$_{n,s}^\textnormal{ non-dispatchable}$ according to the source profile throughout the year. Fossil fuel and dispatchable renewable plants (such as reservoir hydroelectric), instead, can always provide power up until their stated capacity, P$_{n,s}^\textnormal{ dispatchable}$ for node \textit{n} and source \textit{s}.

\begin{equation}
\label{eq:fossil power balance}
    0 \leq \textnormal{P}_{t,n,s} \leq \textnormal{P}_{n,s}^\textnormal{ dispatchable}
\end{equation}

\begin{equation}
\label{eq:ren power balance}
    0 \leq \textnormal{P}_{t,n,s} \leq \textnormal{P}_{n,s}^\textnormal{ non-dispatchable} \cdot \textnormal{a}_{t,n,s}
\end{equation}

P$_{t,n,s}$ = power provided by the source s (can be dispatchable or not) at time \textit{t} in node \textit{n}

a$_{t,n,s}$ = availability of the renewable source \textit{s} at time \textit{t} in node \textit{n}. It can be a number between 0 and 1 (0 being no obtainable energy and 1 being maximum power available).\\

In addition, for dispatchable generators, an efficiency is presented to calculate the amount of resource exploited, according to the following equation. P$_{t,n,s}^\textnormal{ dispatchable}$ in this case is referred only to dispatchable generation units.

\begin{equation}
\label{eq:commodity consumption}
    \textnormal{E}_{t,n,s}^\textnormal{ resource} = \frac{\textnormal{P}_{t,n,s}}{\eta_{n,s}}
\end{equation}

E$_{t,n,s}^\textnormal{ resource}$ = quantity of commodity source employed by resource \textit{s} at time \textit{t} in node \textit{n}.

$\eta_{n,s}$ = efficiency of the generation source \textit{s} in node \textit{n}.\\

Energy can be exchanged between two nodes, passing through transmission lines that have a nominal capacity value. P$_{p}^\textnormal{ powerline}$ represents the maximum exchangeable power in the transmission line \textit{p}, which goes from one node to another. Each powerline \textit{p} has losses related to the exchange of power, taken into account by the efficiency $\eta_{p}^\textnormal{ transmission losses}$. The energy that can pass through powerline \textit{p} at timestep \textit{t} is E$_{t,p}^\textnormal{ exchanged}$.

\begin{equation} \label{eq:powerlines balance}
    \textnormal{E}_{t,p}^\textnormal{ exchanged} \leq \textnormal{P}_{p}^\textnormal{ powerline} \cdot \eta_{p}^\textnormal{ transmission losses} \cdot \Delta t
\end{equation}

$\Delta t$ = timestep (in this case one hour)

\textit{st} = storage technology, from 1 to ST;\\

Finally, storage technologies need to respect constraints and balances for the daily dispatch, bearing in mind also self-discharging. The first crucial restriction is that the storage content SC$_{t,st}$ of the storage technology \textit{st} at the time step \textit{t} can not exceed the maximum storage content SC$_{st}^\textnormal{ nominal}$ for each technology \textit{st}. In this work, Oemof has been set to maintain the same storage content at the beginning and the end of the time frame considered, thus energy stored describes a mean to achieve flexibility in the system.

\begin{equation}
\label{eq:storage content limit}
    \textnormal{SC}_{t,st} \leq \textnormal{SC}_{st}^\textnormal{ nominal}
\end{equation}

Storage units have to observe another limitation, regarding the storage balance: the equation takes into account the power to charge the storage \textit{st} at time \textit{t} in node \textit{n}, P$_{t,n,st}^\textnormal{ charge}$, and the power to discharge it, P$_{t,n,st}^\textnormal{ discharge}$, with their respective efficiencies, $\eta_{st}^\textnormal{ charge}$ and $\eta_{st}^\textnormal{ discharge}$. The storage unit undergoes through a self-discharging process, contemplated with the efficiency $\eta_{st}^\textnormal{ self}$ applied to the stored energy. Eventually, the exchanged power modifies the storage content of the unit from one time step to another.

\begin{equation} 
\label{eq:storage balance}
    (\textnormal{P}_{t,n,st}^\textnormal{ charge} \cdot \eta_{st}^\textnormal{ charge} - \frac{\textnormal{P}_{t,n,st}^\textnormal{ discharge}}{\eta_{st}^\textnormal{ discharge}}) \cdot \Delta t - (\textnormal{SC}_{t,st} - \textnormal{SC}_{t-1,st} ) \cdot \eta_{st}^\textnormal{self} = \textnormal{SC}_{t,st} - \textnormal{SC}_{t-1,st}
\end{equation}

$\textnormal{SC}_{t,st} - \textnormal{SC}_{t-1,st}$ = stored energy at time \textit{t} in storage technology \textit{st}, given by the difference of the storage content from one time step to another.

\subsection{Expansion capacity investment}
\label{subsubsec: expansion capacity investment}

Oemof framework is suitable for investment optimisation analyses: the solver can decide how to invest in order to expand the technologies capacities if this can help to satisfy the energy demand with an overall lower cost. In general, all Oemof components can be expanded introducing the investment option in the code. In this paper the generation technologies that have the possibility to be expanded are rooftop photovoltaic, Utility Scale photovoltaic, on-shore wind and off-shore wind. The chosen RES generation technologies are in line with National Development Plans for Italy \cite{Italy_NC_ONU,PNIEC}, while others are excluded for different reasons: for example nuclear plants can not be added in the electricity mix due to political choices in the country. The expandable storage options are lithium ion batteries, hydrogen tanks, electrolysers and fuel cells. Energy storage is already provided in the current system by Pumped Hydro Storage (PHS), but in the model it cannot be increased since the exploitation limit has been already reached and there is no further space to build this type of power plants, which require a large amount of constructions and affect the environment \cite{hydro_italy}. Li-ion batteries were detected in the electro-chemical storage panorama as the biggest market player \cite{liion_market}, while many studies suggest that hydrogen can be an interesting option for power systems adequacy, since its cost can be reduced by exploiting it as an energy carrier in other hard-to-abate sectors \cite{COLBERTALDO2018592, EGELANDERIKSEN202131963, KHAREL}. The model thus can provides short-term storage with batteries and long term one with hydrogen, offering a comprehensive solution to store electricity produced.

Oemof offers an annuity function that provides the \textit{Energy Periodical Cost} (Ep cost) for each technology, which is the essential variable to choose if an investment is convenient. The Ep cost is calculated according to Eq. \ref{eq:epcost}. Some inputs are needed for each technology in order to compute the ep cost: capital expenditure (capex) in euros per unit of installed capacity, lifetime in years and a Weighted Averaged Cost of Capital (WACC) to consider the credit for the investment.

\begin{equation}\label{eq:epcost}
\begin{split}
    \textnormal{Ep cost} & = \textnormal{oemof.tools.economics.annuity(capex, lifetime, wacc)} \\
    & = \textnormal{capex} \cdot \frac{\textnormal{wacc} \cdot (1+\textnormal{wacc})^\textnormal{lifetime}}{(1+\textnormal{wacc})^{\textnormal{lifetime}-1}}
    \end{split}
\end{equation}

Besides the economic parameters, the solver requires as inputs the existing technology capacity of the technology (which can be \textit{None}) and its maximum potential foreseen for the future year to which the study is referring to. In the expansion capacity optimisation the objective function has been modified, adding also the investment made on the various technologies and minimising the total cost C$^\textnormal{ tot}$.

\begin{equation}
\label{eq:mincosttotal}
\begin{split}
\textnormal{Min C}^\textnormal{ tot} & = \textnormal{Min} \Big ( \textnormal{C}^\textnormal{op} + \sum_{n=1}^{N} (\sum_{s=1}^{S} \textnormal{C}_{s} \cdot \textnormal{P}_{n,s}^\textnormal{added} + \sum_{st=1}^{ST}\textnormal{C}_{st} \cdot \textnormal{E}_{n,st}^\textnormal{added} + \sum_{p=1}^{P} \textnormal{C}_p \cdot \textnormal{P}_{p}^\textnormal{added} \big ) \Big )
\end{split}
\end{equation}

C$_s$ = capital cost for the expansion of the resource \textit{s}

C$_{st}^\textnormal{ added}$ = capital cost for the expansion of storage technology \textit{st}

C$_p$ = capital cost for the expansion of power lines

P$_{n,s}^\textnormal{ added}$ = power capacity added for source \textit{s} in node \textit{n}

E$_{n,st}^\textnormal{ added}$ = storage capacity for storage \textit{st} in node \textit{n}

P$_{p}^\textnormal{ added}$ = power capacity added for power line \textit{p}\\

The output of this function is the total annual cost for operation and optimization of the system. Insights are also given about the amount of power installed for each technology in each node; this is a very helpful information for policy makers, to understand which should be the aim of the incentives to introduce.\\

Oemof framework allows to introduce a constraint on emissions of the system, modifying the main expansion capacity script. To bring about it correctly, it is necessary to bestow emission factors for the polluting commodity sources, in quantity of contaminants per unit of energy supplied. The total amount of emissions, CO$_{2}^\textnormal{total}$ is calculated as follows.

\begin{equation} \label{eq:emissions}
\textnormal{CO}_2^\textnormal{total} = \sum_{t=1}^{T} \sum_{n=1}^{N} \sum_{s=1}^{S} \textnormal{P}_{t,n,s}^\textnormal{ fossil source} \cdot {\textnormal{co}_{2}}_{s}^\textnormal{factor}
\end{equation}

P$_{t,n,s}^\textnormal{ fossil source}$ = thermal power at time \textit{t} in node \textit{n} from the fossil commodity source \textit{s}

${\textnormal{co}_2}_{s}^\textnormal{factor}$ = emission factor specific to the commodity source \textit{s} in [$\frac{\textnormal{ton of CO}_{2}}{\textnormal{MWh}_{\textnormal{th}}}$]
\\
\\
\end{appendices}

\printbibliography

\end{document}